\documentclass[11pt,a4]{article}

\usepackage{amsthm}
\usepackage{amsmath}    % need for subequations
\usepackage{graphicx}
\usepackage{natbib}
\usepackage{multirow}
\usepackage{verbatim}
\usepackage{setspace}
\usepackage{subfigure}
\usepackage{float}

\usepackage{latexsym}
\usepackage{color}
\usepackage{comment}
\usepackage{courier}

\restylefloat{table}

\usepackage{mathptmx}
\usepackage{amssymb}
\usepackage{epstopdf}
\usepackage{comment}
\usepackage{bm}
\usepackage{eucal}
\usepackage{mathrsfs}
\usepackage{graphicx}
\usepackage{booktabs}
\usepackage{color}

\DeclareMathAlphabet{\mathpzc}{OT1}{pzc}{m}{it}
\DeclareMathAlphabet{\mathcalligra}{T1}{calligra}{m}{n}
\include{GH}

\newtheorem{proposition}{Proposition}
\title{\Large \bf A Censored Bayesian Hierarchical Model For Precipitation%\footnote{This article is currently under review at the {\it Journal of the American Statistical Association}}
}
\date{Version of September 15, 2014 (first revision)}
\author{Yang Liu$^{\dagger}$\footnote{Corresponding Author. Email: Yang.Liu@csiro.au},$\>\>\>$ Philip Kokic$^{\dagger}$ and K.Shuvo Bakar $^{\ddagger}$\\
$^{\dagger}$CSIRO, Canberra, Australia.\\
$^{\ddagger}$Yale University, New Haven, USA.}

\begin{document}
\maketitle
\centerline{\footnotesize Under review at {\it Environmetrics}}
\begin{abstract}
Modelling of precipitation, including extremes, is important for hydrological and agricultural applications. Traditionally, because of large sample properties for data over a large threshold value, generalised Pareto (GP) distributions are often used for modelling extreme rainfall. It can be shown that under certain conditions the generalised hyperbolic (GH) distributions can approximate the power law decay of the GP distribution in the tails. Given their flexible form, this raises the possibility that distributions from the GH family serve as a model for the entire rainfall distribution thus avoiding the need to select a threshold. In this paper, we use a flexible censored hierarchical model that leverages the GH distribution to accommodate data subject to heavy tails and an excessive number of zeros. The fitted model allows estimation of probabilities and return periods of the rainfall extremes, and it produces narrower credible intervals in the tails than the traditional GP method. The model not only fits the tails of the rainfall distribution, but fits the whole distribution very well. It also efficiently represents short-term dependencies in the data so it is suitable for evaluating duration over and below thresholds as well as duration of zero rainfall.
\end{abstract}

\section{Introduction}
\label{s:intro}
Currently, the General Circulation Model (GCM) is still the most reliable tool for generating the future climate change scenarios \citep{ye2011method}. But it is widely known that physical models are inadequate for extremes, due to its coarse spatial resolution and the current incomplete understanding of the climate system. \cite{ye2011method} suggest that the poor performance in regional/local precipitation simulation makes it difficult to directly use GCM outputs in climate change impact on extreme precipitation change studies, because extreme precipitation event is most likely a localised phenomenon. For these reasons, statistical models are often considered when modelling rainfall and rainfall extremes.\\
\indent In fact, statistical modelling of rainfall has important applications in many different fields of research including hydrology, agriculture, and environmental sciences. Within each of these fields information is required on a number of spatial and temporal scales. In hydrological applications the frequency and duration of extreme rainfall events over short time periods is very important \citep{Shao2013, Fowler2007,Tetzlaff2005}. For agricultural modelling climate information is required on the short-term variations associated with extreme and non-extreme events through the realistic simulation of observed data \citep{Keating2003, Kokic2013}, as well as intermediate variations associate with seasonal and inter-seasonal variations, and long-term variations due to causes such as climate change. Environmental science applications often require rainfall data with a high degree of spatial resolution \citep{Ashcroft2011,Ashcroft2012}. For these reasons, statistical approaches that can provide information containing these varying temporal and spatial characteristics is valuable.  To address these requirements we need a very flexible statistical approach that can accurately represent upper and lower extremes, skewed data and varying shapes of the rainfall distribution, as well as reliable estimates of the serial dependencies in these data. The objective of this paper is to describe a unified statistical model that moves towards meeting several of these multiple objectives.\\
\indent This paper illustrates the potential advantages of using a censored generalised hyperbolic (GH) model for rainfall modelling. For extreme rainfall modelling it will be compared to the frequently used approach of fitting a generalised Pareto (GP) distribution to data above a threshold. This approach has been widely used for modelling extremes because the GP distribution is the limiting distribution of statistically independent observations lying above a very large threshold \citep{smith1987estimating}. GP distributions have been extensively used as a primary tool for modelling extreme values in a wide range of disciplines. Among others, \cite{van1986generalized}, \cite{coles2003fully}, \cite{li2005statistical} and \cite{yiou2008weather} use the technique to model the tails of rainfall and temperature distributions, \cite{gilli2006application} use the GP distribution to model extreme financial returns over a large threshold, and \cite{finkenstadt2004extreme} discuss the applications of similar methods in insurance, telecommunications and environmental science. The rationale behind considering a GH model is that the GH distribution has an extremely flexible form, so that it can accommodate datasets with heavy tails and skewness. In this paper it is shown that GH distributions can approximate the power law decay of the GP distribution in the tails, as well as achieve a good fit at the centre and shoulders of the distribution. Fitting a GH model does not involve choosing an appropriate threshold which can be a major drawback of GP models because they do not account for the uncertainty associated with the choice of threshold, and potentially information below the threshold is discarded for estimating extremes. This is a significant advantage of modelling with the GH distribution as it allows us to make inferences of not only the rainfall extremes, but also the entire time series. It is also advantageous because the credible intervals (CI) for return periods of rainfall extremes obtained under the GH model have a larger sample contributing to the estimate. \\
\indent We also compare the GH distribution to the Generalised Weibull (GW) distribution \citep{Mudholkar1996}. This distribution, also known as the extended Burr XII distribution, has been used for modelling extreme values of rainfall \citep{Shao2004,Shao2013}. It has also been widely applied to areas such as economics and insurance \citep{maddala1986limited,kleiber2003statistical}. In this paper, it is shown that due to the more flexible shape of the GH disribution, the GH model has the potential to outperform the GW distribution in its ability to fit the entire rainfall distribution.\\
\indent The underlying model we develop for rainfall is an ARMA process with GARCH errors, whose innovations are independent and identically distributed GH random variables. The ARMA and GARCH processes account for the autocorrelation and heteroscedasticity of the time series, whereas the GH distribution addresses the skewed and heavy-tailed nature of rainfall data. Zero rainfall contained in the time series is considered as censored observations from the underlying distribution; an approach that has been employed by various authors to represent dry days in stochastic rainfall models \citep{Hutchinson1995,Wang2012,Sahu2010}. We apply the Metropolis-Hastings algorithm \citep{metropolis} for sampling the joint posterior of model parameters. This enables the easy production of CIs of return periods and other important statistics using Bayesian Markov Chain Monte Carlo (MCMC) sampling.\\
\indent This paper also accounts for various climatic drivers that have influence on precipitation levels. Understanding the effects of climatic drives on rainfall and the pattern of seasonal precipitation is important for agricultural and
water risk-management. Empirical evidences on the relationships between them are discussed in many studies \citep{bakar2013sptimer,shu1,spdy1,Kokic2013}. Three important climatic drivers are the El Nino southern oscillation anomaly (NINO 3.4), Southern Hemisphere Annular Mode (SAMI) and Indian Ocean Dipole (IOD). It has been observed that these indices have potential teleconnection with precipitation levels.\\
\indent This paper is divided into several sections. In the next section, a brief review of the GH, GP and GW distribution is presented. In Section 3, we describe the general setup of the ARMA-GARCH-GH model and model assumptions. In Section 4, we illustrate how to use the censored ARMA-GARCH-GH model for modelling rainfall data. In Section 5 and 6, a simulation study of the ARMA-GARCH-GH model and a sensitivity analysis are presented. In Section 7 we apply the proposed model to rainfall data from three locations (Brisbane, Pardelup and the Oaks), which are chosen in order to test the model in different climatic zones. Finally it will be compared with the GP and GW models. Section 9 includes a short discussion and future directions.

\section[]{background}
\subsection{The Generalised Pareto and Other Related Distributions}
The generalised Pareto (GP) distribution, whose density is given by
\begin{align}
\label{gpl}
d_{GP(\kappa,\mu,\sigma)}(x)=\frac{1}{\sigma}\left(1+\frac{\kappa\left(x-\mu\right)}{\sigma}\right)^{-\left(\frac{1}{\kappa}+1\right)}
\end{align}
is often used for extreme value modelling. This idea was first suggested by \cite{pickands1971two} and has been developed by, among others, \cite{dumouchel1983estimating}, \cite{smith1984threshold}, \cite{smith1987estimating}, \cite{hosking1987parameter} and \cite{joe1987estimation}. In particular, \cite{davison1990models} used a generalised Pareto distribution as the modelling distribution for exceedances over high thresholds. More recently, \cite{li2005statistical} applied this approach, coupled with a threshold value selection criterion suggested by \cite{coles2001introduction}, to model extreme rainfall.\\
\indent Another relevant distribution is the Burr distribution. Since its introduction by \cite{burr1942cumulative}, the Burr distribution has been studied intensively and widely used in many disciplines, particularly economics and actuarial science (\cite{maddala1986limited}, \cite{kleiber2003statistical} and \cite{champernowne1952graduation}). The density function of the Burr distribution is given by
\begin{align}
\label{burr}
f_{Burr\left(c,k,\sigma\right)}(x)=ck\frac{\left(x/\sigma\right)^{c-1}}{\left(1+\left(x/\sigma\right)^c\right)^{k+1}},\>\>\>x>0,\>\>{\text{\normalfont for}}\>c,k,\sigma>0.
\end{align}
We notice that the Burr distribution is in the Pareto family and is a special case of the Pareto type-IV. Thus the Burr distribution is capable of capturing skewness and heavy tails. To further improve its flexibility, the generalised Weibull (GW) distribution (also known as the extended Burr distribution) was introduced by \cite{Mudholkar1996}. The density is given by
\begin{equation}
\label{extburr}
f_{GW\left(c,r,\nu\right)}\left(x\right)=
\left\{
\begin{aligned}
&\frac{c}{\nu}\left(\frac{x}{\nu}\right)^{c-1}\left(1-r\left(\frac{x}{\nu}\right)^c\right)^{\frac{1}{r-1}},\>\>\>r\ne0,\\
&\frac{c}{\nu}\left(\frac{x}{\nu}\right)^{c-1}\exp{\left(-\left(\frac{x}{\nu}\right)^c\right)},\>\>\>r=0.
\end{aligned}
\right.
\end{equation}
We notice that for $r\ne0$, $GW\left(c,r,\nu\right)$ is just the Burr distribution with the reparameterisation $r=-1/k$ and $\nu=\sigma/k^{1/c}$; for $r=0$, \eqref{extburr} reduces to the Weibull distribution. The GW distribution was also used by \cite{Shao2004} and \cite{Shao2013} for extreme flood modelling and dynamic hydrological modelling.

\subsection[]{The Generalised Hyperbolic Distribution}
The generalised hyperbolic (GH) distribution was first introduced by \cite{barndorff1977} in connection with dune movements modelling. Their Lebesgue density is defined as
\begin{align}
\label{newformulation}
 d_{GH(\lambda,\chi,\psi,\mu,\Sigma,\gamma)}(x)=&\frac{\left(\sqrt{\frac{\psi}{\chi}}\right)^\lambda\left(\psi+\gamma^2 \Sigma^{-1}\right)^{\frac{1}{2}-\lambda}}{\sqrt{2\pi\left\vert\Sigma\right\vert}K_\lambda\left(\sqrt{\chi\psi}\right)}\notag\\
&\frac{K_{\lambda-\frac{1}{2}}\left(\sqrt{\left(\chi+\left(x-\mu\right)^2\Sigma^{-1}\right)\left(\psi+\gamma^2\Sigma^{-1}\right)}\right)\exp{\left(\left(x-\mu\right)\Sigma^{-1}\gamma\right)}}{\left(\sqrt{\left(\chi+\left(x-\mu\right)^2\Sigma^{-1}\right)\left(\psi+\gamma^2\Sigma^{-1}\right)}\right)^{\frac{1}{2}-\lambda}},
\end{align}
where $K_v$ is the modified Bessel function of the third kind with index $v$, $\lambda\in\mathbb{R}$, $\chi, \psi\in\mathbb{R}^{+}$ and $\mu, \Sigma, \gamma\in\mathbb{R}$. To gain some intuition on the above expression, one often writes the generalised hyperbolic distribution as the following mean-variance mixture. A random variable $X$ is said to have a GH distribution if
\begin{align}
\label{mvm}
X=\mu+\gamma W+\sqrt{\Sigma}\sqrt{W}Z,
\end{align}
where $Z$ is a Normal random variable with zero mean and unit variance, and $W$ has a generalised inverse Gaussian distribution with parameters $\lambda, \chi$ and $\psi$, or $W\sim GIG(\lambda, \chi, \psi)$, whose density function is given by \eqref{gigden}. Since $X$ given $W=w$ is Normal with conditional mean $\mu+\gamma w$ and variance $\Sigma w$, it is clear that $\mu$ and $\Sigma$ are location and dispersion parameters, respectively. There is a further scale parameter $\chi$, a skewness parameter $\psi$ to allow for flexible tail modelling; and the scalar $\lambda$, which characterises certain subclass and also influences the size of mass contained in the tails.\\
\indent One of the appealing properties of normal mixtures is that the moment generating function of a GH random variable $X$ can be easily calculated using the moment generating function of the generalised inverse Gaussian distribution. In particular, the mean and variance are given by
\begin{align}
\label{mv}
E(X)=\mu+\gamma E(W)\>\>\>{\text{\normalfont and}}\>\>\>Var(X)=\gamma^2 Var(W)+\Sigma E(W).
\end{align}
Another attractive property of the GH distribution is that, as the name suggests, it is of a very general form, and contains as special cases many of important distribution widely used in the literature. It includes, among others, the Student's t-distribution, the Laplace distribution, the hyperbolic distribution, the normal-inverse Gaussian distribution and the variance-gamma distribution. It is often used in economics, with particular application in the fields of modelling financial markets and risk management, due to its semi-heavy tails.\\
\indent \cite{Shao2004} has shown that the GW distribution can approximate the GP distribution in the tails. We now present a similar result for the GH distribution.
\begin{proposition}
\label{thm1}
For any fixed threshold $\mu$, the power law decay of the GP distribution, whose density function is given by \eqref{gpl} can be approximated by the GH distribution in the tails if the shape parameter $\kappa$ of the GP distribution satisfies either of the two conditions:
\begin{enumerate}
\item $\kappa>0$ (heavy tail);
\item $\kappa\to0$ (light tail).
\end{enumerate}
\end{proposition}
The proof is given in Appendix \ref{app1}. We note that the tail for a GP distribution cannot be approximated by the GH distribution if the GP distribution has a bounded tail, i.e. $\kappa<0$. However this scenario is unrealistic for precipitation data. Therefore Proposition \ref{thm1} suggests the possibility that distributions from the GH family serve as the underlying distribution for rainfall data.\\
\indent There are several parameterisations of the GH distribution. The $(\lambda,\chi,\psi,\mu,\Sigma,\gamma)$ parameterisation used in this paper has a drawback of an identification problem \citep{barndorff2001normal}, i.e. $GH(\lambda,\chi,\psi,\mu,\Sigma,\gamma)$ and $GH(\lambda,\chi/k,k\psi,\mu,k\Sigma,k\gamma)$ are identical for any $k>0$. This is because that $\chi$ and $\Sigma$ are not separately identified. Therefore, an identification problem can be an issue when we try to fit the GH distribution to data. This problem can be solved by introducing a suitable constraint on the parameters. \cite{barndorff2001normal} fixed the dispersion parameter $\Sigma$ to be 1, or in the multivariate case, the determinant of $\Sigma$ to be 1. However under their setup, it is difficult to reparameterise the GH distribution so that it has mean zero and unit variance because this involves reparameterising scalars inside convolutions of Bessel functions with different indices. Such standardisation is necessary when we develop the ARMA-GARCH-GH model in the next section. Fortunately, \cite{JEd} suggested that the identification problem can be also solved by fixing $\chi=1$. We will use their result here, because it removes the need for reparameterising the parameter $\chi$ inside the Bessel function, hence leads to a simple standardisation (see Appendix \ref{app2} for proof).
\begin{proposition}
\label{prop1}
Let us modify \eqref{newformulation} by replacing $\gamma=\tau\Sigma$, $\tau>0$, we have $X\sim GH(\lambda,\chi,\psi,\mu,\Sigma,\tau)$. For any $\lambda, \psi, \tau$ and fixed $\chi=1$, if $\mu$ and $\Sigma$ satisfy
\begin{align}
\label{muvar}
\mu=-\frac{\tau\Sigma M_{\lambda+1}\left(\sqrt{\psi}\right)}{\sqrt{\psi}}\>\>\>{\text{\normalfont and}}\>\>\>\Sigma=\frac{\sqrt{1+4\left(N_{\lambda+2}\left(\sqrt{\psi}\right)-1\right)\tau^2}-1}{2\tau^2M_{\lambda+1}\left(\sqrt{\psi}\right)\left(N_{\lambda+2}\left(\sqrt{\psi}\right)-1\right)/\sqrt{\psi}},
\end{align}
where
\begin{align}
M_{\lambda+1}\left(\sqrt{\psi}\right)=\frac{K_{\lambda+1}\left(\sqrt{\psi}\right)}{K_{\lambda}\left(\sqrt{\psi}\right)}\>\>\>{\text{\normalfont and}}\>\>\>N_{\lambda+2}\left(\sqrt{\psi}\right)=\frac{K_{\lambda+2}\left(\sqrt{\psi}\right)K_{\lambda}\left(\sqrt{\psi}\right)}{K_{\lambda+1}\left(\sqrt{\psi}\right)^2},\notag
\end{align}
then $E(X)=0$ and $Var(X)=1$.
\end{proposition}
We note that without modifying \eqref{newformulation} as in Proposition \ref{prop1}, one could simply write $\mu$ and $\Sigma$ as expressions of $\lambda, \psi$ and $\gamma$, but this does not guarantee the positivity of $\Sigma$. This problem can be solved by introducing the parameter $\tau$ to create a quadratic function of $\Sigma$. After standardisation the original $(\lambda,\chi,\psi,\mu,\Sigma,\gamma)$ parameterisation is used to make use of the {\bf R} package \texttt{ghyp} \citep{breymann2013ghyp} for simulation.\\
\indent Unlike models based on the extreme value theory, such as the GP distribution, which disregards data below a large threshold and does not account for the uncertainty associated with the choice of threshold, fitting a GH model does not involve choosing such a threshold. This is a significant advantage of modelling with the GH distribution as it allows us to make inferences of not only the rainfall extremes, but also the entire time series. Consequently the prediction credible intervals (CI) for return periods of rainfall extremes and other statistical inference obtained under the GH model have a larger sample contributing to the estimates. We will now construct our Bayesian hierarchical model and demonstrate these advantages.

%%%%%%%%%%%%%%%%%%%%%%%%%%%%%%%%%%%%%%%%%%%%%%%%%%%%%%%%%%%%%%%%%%

\section{ARMA-GARCH-GH Model}
\label{ss:example}
In this section, a formulation of the proposed statistical model for the underlying distribution of rainfall is presented. Before we introduce the model, there are four important features of rainfall data that need to be addressed, namely autocorrelation, heteroscedasticity, heavy-tailedness and an excessive amount of zero observations. We will discuss the first three here and deal with the zero problem in Section \ref{data}.\\
\indent Let us consider historical rainfall data from the Brisbane climate observation station in south-east Queensland obtained from the Commonwealth Bureau of Meteorology. The sample period is from 1 January 1889, to 22 January 2014, for a total of 45,677 observations, including 22,677 wet season (November--April) observations. The time series of the wet season daily rainfall is shown in Figure 1. The middle panel shows a plot of the autocorrelation function, and indicates that there is autocorrelation present in the data. The right panel shows rainfall volatility during the period of record, from which strong heteroscedasticity can be identified. This may be a consequence of seasonal variation and alternating dry-wet changes. The heavy-tailed nature of the rainfall distribution is illustrated in both the left and right panels, where the majority of observations are scattered below 75 mm, many others distributed between 100 and 150 mm, and a few days with heavy rainfall over 200 mm are observed, which are considered rare or extreme events and are of great interest.\\
\begin{figure}
    \centerline{
    \includegraphics[width=160mm]{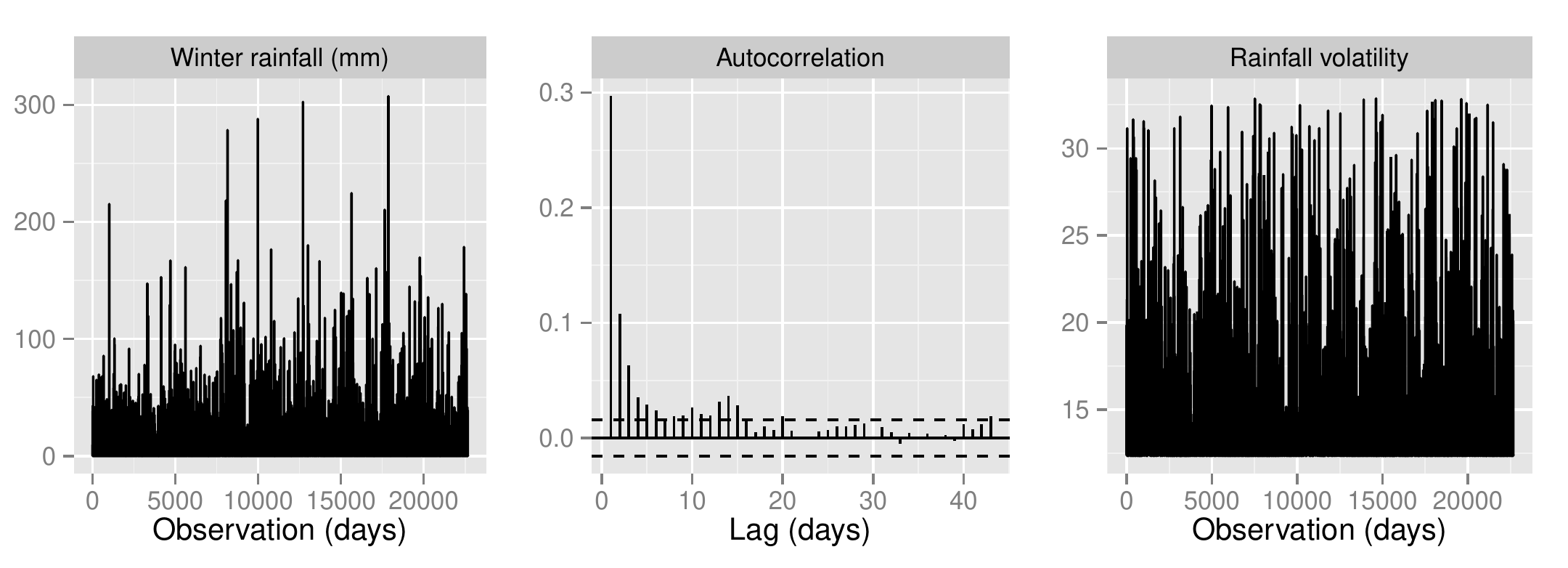}}
    \caption{\footnotesize Rainfall data from 1 January 1889, to 22 January 2014 at the weather station Brisbane, Queensland. The left panel is the time series of summer daily rainfall (mm); the middle panel shows a plot of the autocorrelation function, and indicates that there is autocorrelation present in the data; the right panel shows rainfall volatility in the period of record, from which strong heteroscedasticity can be identified. Thus there are sub-samples that have different variabilities from others.
}
    \label{Rplot40}
\end{figure}
\indent To address the autocorrelation, heteroscedasticity and heavy-tailed features of rainfall data, we propose the following model. An autoregressive and moving average (ARMA, \cite{arma}) model is used to remove persistence. It generates rainfall residuals with no correlation with past values. A generalised autoregressive conditional heteroscedasticity (GARCH, \cite{bollerslev1986generalized}) model is also used to account for empirical features in the volatility of rainfall observations. Finally the skewness and heavy-tailed nature of the rainfall distribution can be modelled with a standardised GH distribution. We are now in a position to introduce the ARMA-GARCH-GH model:
\begin{align}
\label{m1}
&y_t=a_0+\sum^{p}_{i=1}a_iy_{t-i}+\sum^{q}_{i=1}b_i\epsilon_{t-i}+\sum^M_{i=1}\beta_i u_{it}+\epsilon_t\\
\label{m2}
&\epsilon_t=z_t\sigma_t,\\
\label{m3}
&\sigma^2_t=\alpha_0+\sum^{m}_{i=1}\alpha_i\epsilon_{t-i}^2+\sum^{r}_{i=1}\omega_i\sigma^2_{t-i},
\end{align}
where $y_t$, $t=1,\dots,T$ is the observed data, $u_i$ is the $i$th covariate with coefficient $\beta_i$, and $z_t\sim GH(\lambda,\chi,\psi,\mu,\Sigma,\gamma)$ are independent and identically distributed standardised GH random variables with zero mean and unit variance. We let $\Phi\left(L\right)=\sum^p_{i=1}a_iL^i$, $\Theta\left(L\right)=\sum^q_{i=1}b_iL^i$, $A\left(L\right)=\sum^m_{i=1}\alpha_iL^i$ and $B\left(L\right)=\sum^r_{i=1}\omega_{j}L^j$, where $L$ is the lag operator.\\
\indent The mean process is given by \eqref{m1} with autoregressive coefficients $a_i$ for $i=1,\dots,p$ and moving average coefficients $b_i$ for $i=1,\dots,q$. In a standard ARMA model, however, the conditional variance given the past is constant, so it cannot take into account for heteroscedasticity of the time series. In this context we use the GARCH model given by \eqref{m2} and \eqref{m3} in which the conditional variance of innovations $\epsilon_t$ is non constant and depends on time, and therefore can model the randomly varying volatility.\\
\indent  In practice, we impose several constraints on the parameters in the ARMA-GARCH-GH model:
\begin{enumerate}
\item All roots of $1-\Phi\left(L\right)=0$ and $1+\Theta\left(L\right)=0$ are outside the unit circle;
\label{c2}
\item $\alpha_0>0$, $\alpha_i>0$ for $i=1,\dots,m$ and $\omega_i>0$ for $i=1,\dots,r$;
\label{c4}
\item permitted parameters of GH distributions are $\lambda\in\mathbb{R}$, $\chi, \psi\in\mathbb{R}^{+}$ and $\mu, \Sigma, \gamma\in\mathbb{R}$. Using Proposition \ref{prop1} the GH distribution can be standardised so that it has zero mean and unit variance, so $\Sigma$ and $\mu$ can be written as functions of $\lambda$, $\psi$ and $\gamma$. This requires $\chi=1$.
\end{enumerate}
Condition \ref{c2} is to ensure the stationarity and invertibility of the $ARMA(p,q)$ process; \ref{c4} is imposed to guarantee that the conditional variance $\sigma^2_t$ is always positive. \cite{bollerslev1986generalized} shows that the $GARCH(m,r)$ process is weak-sense stationary if and only if $A(1)+B(1)<1$, and it has mean and variance $E\left(\epsilon_t\right)=0$ and $Var\left(\epsilon_t\right)=\alpha_0/\left(1-A(1)-B(1)\right)$. However, we will not impose the stationarity constraint on the GARCH model since the Bayesian analysis enables us to test variance stationarity condition and estimate the density of the unconditional variance $Var\left(\epsilon_t\right)$ when the condition is satisfied.\\

%%%%%%%%%%%%%%%%%%%%%%%%%%%%%%%%%%%%%%%%%%%%%%%%%%%%%%%%%%%%%%%%%%

\section{Censored Bayesian Hierarchical Model For Precipitation}
\label{data}
In this section, we describe an application of the above argument and illustrate how to apply the ARMA-GARCH-GH model incorporated with covariates to rainfall data. The rainfall data shows strong autocorrelation and heteroscedasticity that must be adjusted for. Another important feature of the dataset is that it contains an excessive number of zeros, which are considered as censored observations from the underlying process given by \eqref{m1}, \eqref{m2} and \eqref{m3}.\\
\indent We begin by describing weekly rainfall sums $X_t=\left(x_1,\dots,x_T\right)$ by the following censored process. Here we only consider $ARMA\left(1,0\right)$ and $GARCH\left(0,1\right)$, higher order models follow by reproducing carefully the result, keeping track of the time dependence:
\begin{align}
&x_t=y_tI_{\left\{y_t>0\right\}}\\
&y_t=a_0+a_1y_{t-1}+\sum^M_{i=1}\beta_i u_{it}+\epsilon_t\\
&\epsilon_t=z_t\sigma_t\\
&\sigma^2_t=\alpha_0+\alpha_1\epsilon_{t-1}^2,
\end{align}
where $u_{it}$ is the $i$th regressor with coefficient $\beta_i$, $I_{\left\{\cdot\right\}}$ is the indicator function, and $z_t\sim GH\left(\lambda,\chi,\psi,\mu,\Sigma,\gamma\right), \gamma=\tau\Sigma$ are independent and identically distributed with zero mean and unit variance. \\
\indent Let us first focus on the uncensored model. Let $\theta=\left(\lambda,\chi,\psi,\mu,\Sigma,\tau\right)$, $\varphi=\left(a_0,a_1,\beta_1,\dots,\beta_M,\alpha_0,\alpha_1,\theta\right)$ and $\varphi^*=\left(\varphi,y_0,\epsilon_0,\sigma_0\right)$, where $\sigma_0=\sqrt{\alpha_0}$, $y_0=\epsilon_0$ and $\epsilon_0$ is the pre sample error. We also assume $\epsilon_i=\sigma_i=y_i=0$ for $i<0$. To perform Bayesian analysis for the ARMA-GARCH-GH model, we construct the posterior density function of the model:
\begin{align}
\label{JP1}
\pi\left(\varphi^*\vert Y\right)\varpropto f\left(Y\vert\varphi^*\right)\pi(\varphi^*),
\end{align}
where $f\left(Y\vert\varphi^*\right)$ is the likelihood function and $\pi(\varphi^*)$ is the prior. The likelihood function has the following representation:
\begin{align}
\label{like1}
f\left(Y\vert\varphi^*\right)=&f\left(y_1\vert\varphi^*,\dots,y_{-1},y_0\right)f\left(y_2\vert\varphi^*,\dots,y_{-1},y_0,y_1\right)f\left(y_3\vert\varphi^*,\dots,y_{-1},y_{0},y_1,y_2\right),\dots,\notag\\
&f\left(y_T\vert\varphi^*,\dots,y_{-1},y_0,y_1,\dots,y_{T-1}\right).
\end{align}
For any term in \eqref{like1}, say $f\left(y_t\vert\varphi^*\dots,y_{-1},y_0,y_1,\dots,y_{t-1}\right)$, we know
\begin{align}
\frac{y_t-\mu_t}{\sigma_t}=z_t\sim GH\left(\lambda,\chi,\psi,\mu,\Sigma,\gamma\right),\>\gamma=\beta\Sigma,
\end{align}
then $y_t\sim GH\left(\lambda,\chi,\psi,\mu\sigma_t+\mu_t,\Sigma\sigma_t^2,\gamma\sigma_t\right)$, where
\begin{align}
\mu_t=a_0+a_1y_{t-1}+\sum^M_{i=1}\beta_i u_{it}\>\>\>{\text{\normalfont and}}\>\>\>\sigma_t=\sqrt{\alpha_0+\alpha_1\epsilon_{t-1}^2}.\notag
\end{align}
If we write $\widetilde{\Sigma_t}=\Sigma\sigma_t^2$, $\widetilde{\gamma_t}=\gamma\sigma_t$ and $\widetilde{\mu_t}=\mu\sigma_t+\mu_t$, the joint posterior is given by
\begin{align}
\label{cony}
f\left(\varphi^*\vert Y\right)=&\prod^{T}_{t=1}\frac{\left(\sqrt{\frac{\psi}{\chi}}\right)^\lambda\left(\psi+\widetilde{\gamma_t}^2 \widetilde{\Sigma_t}^{-1}\right)^{\frac{1}{2}-\lambda}}{\sqrt{2\pi\left\vert\widetilde{\Sigma_t}\right\vert}K_\lambda\left(\sqrt{\chi\psi}\right)}\times\notag\\
&\prod^{T}_{t=1}\frac{K_{\lambda-\frac{1}{2}}\left(\sqrt{\left(\chi+\left(x-\widetilde{\mu_t}\right)^2\right)\left(\psi+\widetilde{\gamma_t}^2\widetilde{\Sigma_t}^{-1}\right)}\right)\exp{\left(\left(x-\widetilde{\mu_t}\right)\widetilde{\Sigma_t}^{-1}\widetilde{\gamma_t}\right)}}{\left(\sqrt{\left(\chi+\left(x-\widetilde{\mu_t}\right)^2\right)\left(\psi+\widetilde{\gamma_t}^2\widetilde{\Sigma_t}^{-1}\right)}\right)^{\frac{1}{2}-\lambda}}\times\notag\\
&\pi\left(a_0\right)\pi\left(a_1\right)\prod^M_{i=1}\pi\left(\beta_i\right)\pi\left(\alpha_0\right)\pi\left(\alpha_1\right)\pi\left(\theta\right).
\end{align}
We use normal priors on the parameters of the mean process, truncated normal priors on GARCH parameters, and normal and truncated normal priors on GH parameters.\\
\indent Due to the recursive nature of ARMA and GARCH models, the joint posterior and full conditional densities cannot be expressed in closed form. Therefore we cannot use the Gibbs sampler and need to rely on a more elaborated Metropolis-Hasting (MH) simulation strategy to approximate the posterior density \citep{metropolis,hastings}. The sampling strategy relies on the construction of a Markov chain with realisations $\varphi^{[0]},\varphi^{[1]},\dots,\varphi^{[j]},\dots$ in the parameter space. Under appropriate regularity conditions, asymptotic results guarantee that as $j\to\infty$, $\varphi^{[j]}$ converges in distribution to a random variable whose density is $\pi\left(\varphi\vert Y\right)$. Hence after discarding a burn-in of the first draws, the realised values of the chain can be used to make inference about the joint posterior. In this paper, a modified Metropolis-Hastings algorithm was used for the sampling the joint posterior. We use normal proposals and whose variances are tuned to give an acceptance rate around 30\%. The block sampling techique is also used to address the correlation between the GH parameters.\\
\indent The above algorithm only applies when all $y_t$ are observed, that is, $x_t=y_t$ for all $t$. Due to the zero-inflated nature of the dataset, however, we are not able to observe the underlying process $Y_t$ directly for all t. Instead we observe $Y_t$ only when it is positive, so $X_t$ is the value we observe due to censoring. We use the conditional density of $y_t$ given by \eqref{cony} to simulate censored observations and recover the underlying process.\\
\indent Suppose $y_1,\dots,y_{t-1}$ are observed and $y_t$ is censored, i.e. $x_i=y_i$ for $i=1,\dots,t-1$, and $x_{t}=0$, and let $\varphi^{[j]}=\left(a_0^{[j]},a_1^{[j]}, \beta_1^{[j]},\dots,\beta_M^{[j]}, \alpha_0^{[j]}, \alpha_1^{[j]}, \lambda^{[j]}, \psi^{[j]},\tau^{[j]}\right)$ be the $j^{th}$ realisation of the Markov chain in the parameter space. To recover the uncensored process $Y_t$ at time $t$, we sample from the distribution of $Y_t$ conditional on $Y_{t} < 0$ and given the values $y_{t-1},y_{t-2}\dots$, where
\begin{align}
\label{tun}
&Y_t\sim GH_{\left(-\infty,0\right]}\left(\lambda^{[j]},1,\psi^{[j]},\mu^{[j]}\sigma_t+\mu_t,\Sigma^{[j]}\sigma_t^2,\gamma^{[j]}\sigma_t\right),\\
&\Sigma^{[j]}=\frac{\sqrt{1+4\left(N_{\lambda^{[j]}+1}\left(\sqrt{\psi^{[j]}}\right)-1\right)\left(\tau^{[j]}\right)^2}-1}{2\left(\tau^{[j]}\right)^2M_{\lambda^{[j]}+1}\left(\sqrt{\psi^{[j]}}\right)\left(N_{\lambda^{[j]}+1}\left(\sqrt{\psi^{[j]}}\right)-1\right)/\sqrt{\psi^{[j]}}},\notag\\
&\mu^{[j]}=-\frac{\tau^{[j]}\Sigma^{[j]} M_{\lambda^{[j]}+1}\left(\sqrt{\psi^{[j]}}\right)}{\sqrt{\psi^{[j]}}},\notag\\
&\gamma^{[j]}=\tau^{[j]}\Sigma^{[j]},\notag\\
&\sigma_t=\sqrt{\alpha_0^{[j]}+\alpha_1^{[j]}\epsilon_{t-1}^2},\notag\\
&\mu_t=a_0^{[j]}+a_1^{[j]}x_{t-1}+\sum^M_{i=1}\beta_i^{[j]} u_{it},\notag
\end{align}
and $ GH_{\left(-\infty,0\right]}$ is the normalised truncated GH distribution to $ {\left(-\infty,0\right]}$. For later censored observations $x_{t_k}$, $t_k>t$, the simulated $Y_t$ is used as an observed value. The data augmentation process is repeated for each MCMC iteration. Therefore, by simulation we are able to recover the underlying process $y_t$ for all $t$, and for all sets of parameter realisations $\varphi^{[j]}$ in the MCMC chain. This approach to recovering the latent process is quite standard in models involving censoring, see e.g. \cite{Sahu2010}.\\
\indent We produce CIs for return periods by independently simulating time series a large number of times using the MH algorithm at each MCMC iteration and \eqref{tun} from the ARMA-GARCH-GH model, and then by selecting the 2.5\% and 97.5\% quantiles from the estimated return periods.

\section{Simulation Study}
In this section we present a short simulation study for the proposed ARMA-GARCH-GH model. The model is applied to simulated data and the parameter estimates are compared to their true values. The simulated data is comparable to the real rainfall data and it contains 3,200 observations with a censoring rate of around 8\%. The simulation result is shown in Table 1. It is evident that the 95\% CIs for the parameters contain their true values.
\begin{table}[ht] 
\centering % used for centering table 
\begin{tabular}{cccc} % centered columns (4 columns) 
\hline\hline %inserts double horizontal lines 
Parameters & True values& Posterior mean &  95\% CIs \\ [0.5ex] 
%heading 
\hline % inserts single horizontal line 
$a_0$ & 5 & 4.91&4.82, 5.09\\ 
$a_1$ & 0.5 & 0.51&0.48, 0.54 \\ 
$\alpha_0$ & 13 & 13.36&12.50, 14.89 \\ 
$\alpha_1$ & 0.2 & 0.19&0.16, 0.23 \\ 
$\lambda$ & -0.2 & -0.17&-0.60, 0.02 \\ 
$\tau$ & 15 & 15.71&9.30, 25.18 \\ 
$\psi$ & 0.25 & 0.27&0.16, 0.42\\ 
\hline \\[1ex] 
\end{tabular} 
\label{table:nonlin} % is used to refer this table in the text 
\caption{A simulation study of the ARMA-GARCH-GH model.} % title of Table 
\end{table} 
\section{Sensitivity Analysis}
One of the disadvantages of Bayesian analysis is due to the fact that the prior distribution for the parameters can have a significant impact on the posterior distribution, and consequently, leads to biased results. We check the sensitivity of the model by using diffierent  hyper-parameters of the Normal and truncated Normal priors. In the case where the original priors are uniform distributions, Beta distributions on the same support are used as alternatives. The results show that our model is not very sensitive to the choice of the hyper-parameter values. For brevity these results have been omitted from the paper.
\section{Modelling Results}
We apply the ARMA-GARCH-GH model to rainfall data from three locations: Pardelup in south-west Western Australia, The Oaks in south-east New South Wales and Brisbane in south-east Queensland. These locations were chosen in order to test the model in different climatic situations. The first location is in a temperate winter-rainfall dominated zone with a greater number of extreme rainfall events having been observed in recent decades. The second location is on the SW edge of the Sydney basin, and being located close to the coast it experiences a relatively wet temperate climate. However, the area's weather is highly variable: drought and bushfire on the one hand, and storms and heavy rainfall on the other. The last location is a metropolis with a population of more than two million, and in a summer rainfall dominated sub-tropical climate zone. During summer, thunderstorms are common over Brisbane, with the more severe events accompanied by large damaging torrential rain and destructive winds. The city also lies in the tropical cyclone risk area. The last to affect Brisbane was Tropical Cyclone Hamish in March 2009. \\
\indent Three important climatic drivers the El Ni˜no southern oscillation anomaly (NINO 3.4), Southern Hemisphere Annular Mode (SAMI) and Indian Ocean Dipole (IOD) are considered in this study to understand their effects on precipitation levels. The NINO 3.4 index is a monthly time series of mean sea surface temperature (SST) index from the El Nino region that covers 5 south to 5 north and 125 west to 175 east. The SAMI index includes the Antarctic oscillation and obtained by the differences in the normalised monthly zonal mean seal level pressure between 40 and 70 south. The IOD index used in this study is calculated from monthly SST anomaly, which is a coupled ocean and atmosphere phenomenon in the equatorial Indian Ocean. In addition, for weather station Brisbane we also consider the Norfolk-Hawaii index (NHI). The NHI is an index of the Inter-decadal Pacific Oscillation (IPO). The IPO is a slow background change in Pacific Ocean SSTs, which affects the relationship between the El Nino-Southern Oscillation and Queensland summer rainfall.
\subsection{Model Based Analysis}
We present the results for three locations below and discuss in detail the result for Brisbane, whereas results for Pardelup and the Oaks is presented in Appendix B. The models are fitted with 10,000 MCMC iterations, and the first 3,000 samples are discarded as burn-in. The MCMC chains converged quickly after a few hundred iterations. For brevity, these results and other MCMC diagnostics are omitted. \\
\indent Table 2 shows the model parameter estimates and their 95\% CIs. The autoregressive parameters show a significant positive autocorrelation between successive weeks for rainfall at the Oaks and Brisbane. The GARCH parameter $\alpha_1$ is significant at all locations, indicating that there are a considerable amount of clustering in weekly rainfall data. As expected, the generalised hyperbolic shape and skewness parameters show that the estimated distribution is highly skewed to the right. A negative and significant coefficient for NINO3.4 and SAMI at the Oaks is consistant with their effect as reported by the Australian Bureau of Meteorology (http://www.bom.gov.au). The sign of the IPO coefficient is consistent with the asymmetric response identified by \cite{king2013asymmetry}.\\
\begin{table}
\centering
\begin{tabular}{lllllll}
\toprule 
\toprule 
    Parameters & \multicolumn{2}{c}{The Oaks} & \multicolumn{2}{c}{Brisbane}& \multicolumn{2}{c}{Pardelup}\\
\toprule 
    & mean & 95\% CIs &  mean & 95\% CIs&  mean & 95\% CIs \\ 
    \midrule
$a_0$ &11.42 & 10.55,12.40& 30.20&29.59,30.89 & 20.62& 18.30,22.65 \\ 
$a_1$ & 0.07 & 0.01,0.13& 0.17 &0.01,0.35& 0.04& -0.04, 0.14\\ 
$\alpha_0$ & 152.26 & 115.38,204.64  & 2508.63 & 2296.72,2914.88 & 271.24& 209.20,427.56\\ 
$\alpha_1$ & 0.47 & 0.2558,0.77  &0.67 &0.05,2.27& 0.11& 0.01,0.28\\ 
$\lambda$ &0.27&0.04,0.59  & 0.48 &0.39,0.57 & -1.86& -3.08,-0.59\\ 
$\beta$ & 0.49& 0.33,0.68 & 1.36 & 1.15,1.61 & 0.51 &0.24,0.79\\ 
$\psi$ & 0.20 & 0.01,0.49 & 0.07$\times 10^{-2}$ &0.00, 0.28$\times 10^{-2}$ & 0.82& 0.01,2.8\\ 
$\beta1$ (NINO3.4) & -0.08 & -0.24,0.00 & -0.45  &-1.11,0.41& -0.76& -3.08,1.07\\ 
$\beta2$ (SAMI) &-0.50 &-0.92,-0.09& - & - & -0.37& -0.96,0.21\\ 
$\beta3$ (IOD) &-0.03 &-0.88,0.84 & - &-& -0.74& -1.88,0.39\\ 
$\beta4$ (IPO) & - & - & 0.96 & 0.09,2.33 &-& -\\ 
    \bottomrule
\end{tabular}
 \label{restab} % is used to refer this table in the text 
\caption{Posterior mean and corresponding 95\% CIs of the parameters of the ARMA-GARCH-GH model.} % title of Table 
\end{table} 
\indent We should also recognise that the lagged effects of climatic drivers on precipitations can be important \citep{drosdowsky2001near, stone1996prediction}. \cite{schepen2012evidence} argued that concurrent relationships between climatic drivers and precipitations does not imply a lagged relationship, so it is important to understand the effects of lagged climate indices on rainfall. However it is outside our scope, and these exercises have been omitted from the paper.\\
\indent We can test the variance stationarity condition and estimate the unconditional variance of $\epsilon_t$ when the condition is met. As shown by \cite{bollerslev1986generalized}, under the $GARCH(1,1)$ specification, the process is stationary if $\alpha_1+\beta_1<1$. With a value close to one, past volatility will have a longer impact on the future conditional variance. Under $GARCH(0,1)$, the posterior density of $\alpha_1$ has a mean of $0.47$, $0.57$ and $0.11$ at three locations. A simple t-test shows that the model satisfies the variance stationarity condition. The density of the unconditional variance $Var\left(\epsilon_t\right)$ can be estimated using $\alpha_0/\left(1-\alpha_1\right)$ if required.
\begin{figure}[t]
 \centerline{\includegraphics[width=175mm]{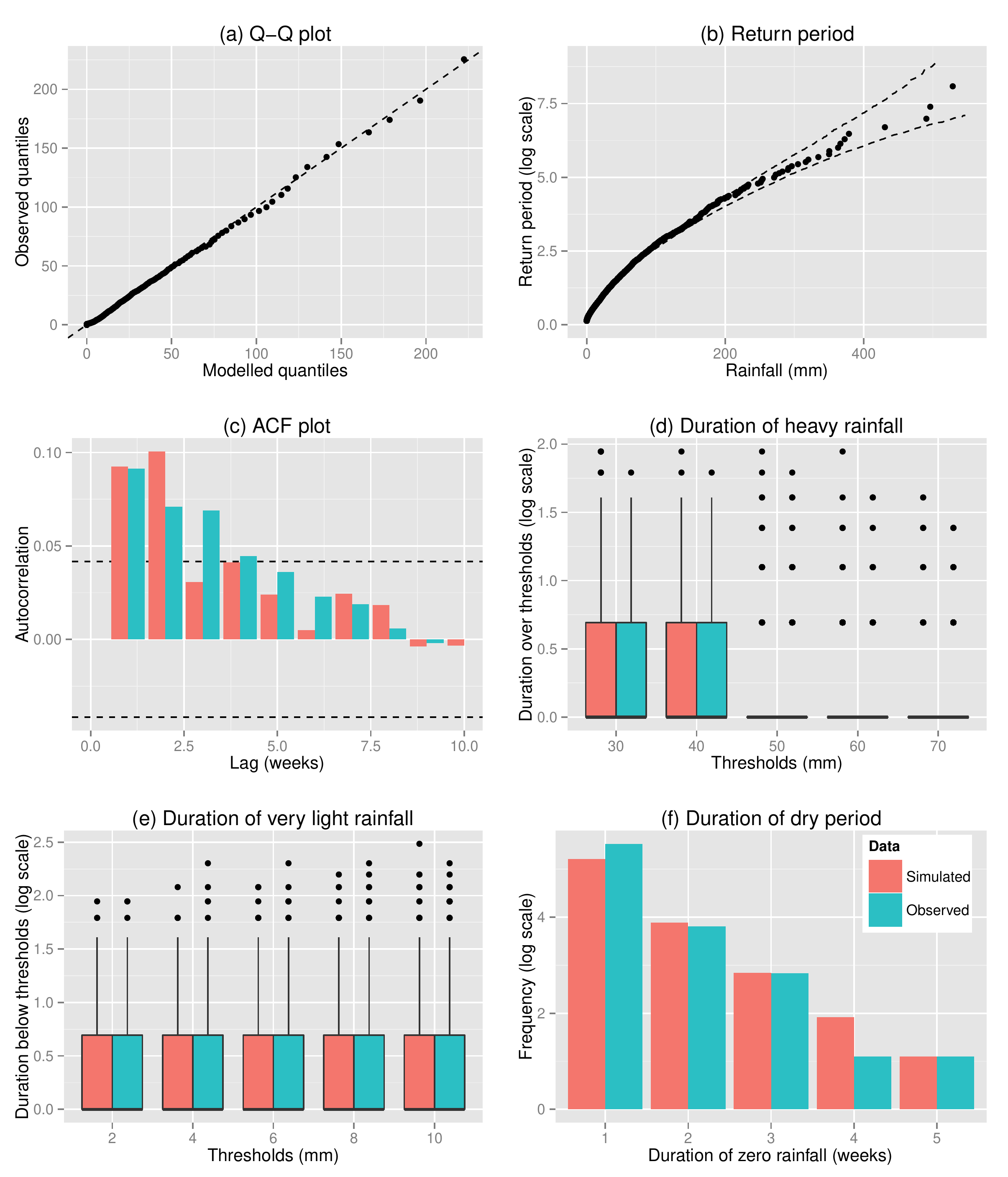}}
 \caption{\footnotesize Modelling results for the weather station Brisbane. Panel (a) is a Q-Q plot where empirical quantiles are plotted on the y-axis, and modelled quantiles are on the x-axis. Panel (b) shows return periods of the wet season weekly rainfall. The dots are empirical return periods, and the 95\% CI given as dashed lines. Panel (c) shows the autocorrelation function of the data simulated by the model comparing to the observed autocorrelation function of the rainfall data. Panel (d) shows the distributions of numbers of weeks with simulated rainfall over chosen thresholds, comparing to the empirical distributions. Panel (e) is a plot of duration of very light rainfall, showing the distributions of numbers of consecutive weeks with rainfall less than certain amount, also comparing to the empirical distribution. Finally Panel (f) shows durations of zero rainfall for both observed and simulated data. All based on 1,000 simulations.}
\label{acf}
\end{figure}
\subsection{Model validations}
Modelling validation for the Brisbane weather station are shown in Figure 2. Panel (a) is a Q-Q plot where empirical quantiles are plotted on the y-axis, and modelled quantiles are on the x-axis. The empirical return periods of the wet season extreme weekly rainfall in Brisbane is presented in Panel (b) with the 95\% CI given as dashed lines. We see that the ARMA-GARCH-GH model not only fits the tails of the rainfall distribution  but fits the whole distribution very well. Furthermore, the model captures the dependence among rainfall observations. Panel (c) plots the autocorrelation function of the data simulated by the ARMA-GARCH-GH model against the observed autocorrelation function of the rainfall data. We notice that the autocorrelation of two datasets almost align for small lags. Since the model can efficiently represent short-term dependencies in the data, it is suitable for evaluating duration over thresholds. Panel (d) shows that the model produces a good estimate of the distributions of numbers of weeks with rainfall over chosen thresholds, comparing to the empirical distributions. Panel (e) is a plot of duration of very light rainfall, showing the distributions of numbers of consecutive weeks with rainfall less than certain amount. Finally Panel (f) shows durations of zero rainfall. It illustrates the numbers of consecutive weeks with no rain as a bar plot. Overall the ARMA-GARCH-GH model fits the rainfall data well.

\section[]{Comparison to the Generalised Pareto and Weibull Distribution}
We compare our censored ARMA-GARCH-GH method to the widely used approach based on a generalised Pareto distribution. Using the method adopted by \cite{li2005statistical}, we choose thresholds $u=150, 200$ and $250$ for the GP distribution. We compare the results of the Pareto model with the ARMA-GARCH-GH model in the top panel of Figure 3, both applied to the Brisbane weather station. It shows that the CI estimates vary when the threshold changes, but the GH model always produces narrower CI in the tails. This indicates that useful information from the centre of the data is being discarded for estimating extreme rainfall return periods by the GP approach.   
\begin{figure}
\centering
 \begin{minipage}[b]{0.75\linewidth}
    \centering
   \centerline{\includegraphics[trim = 0mm 0mm 0mm 6mm, clip,width=\textwidth]{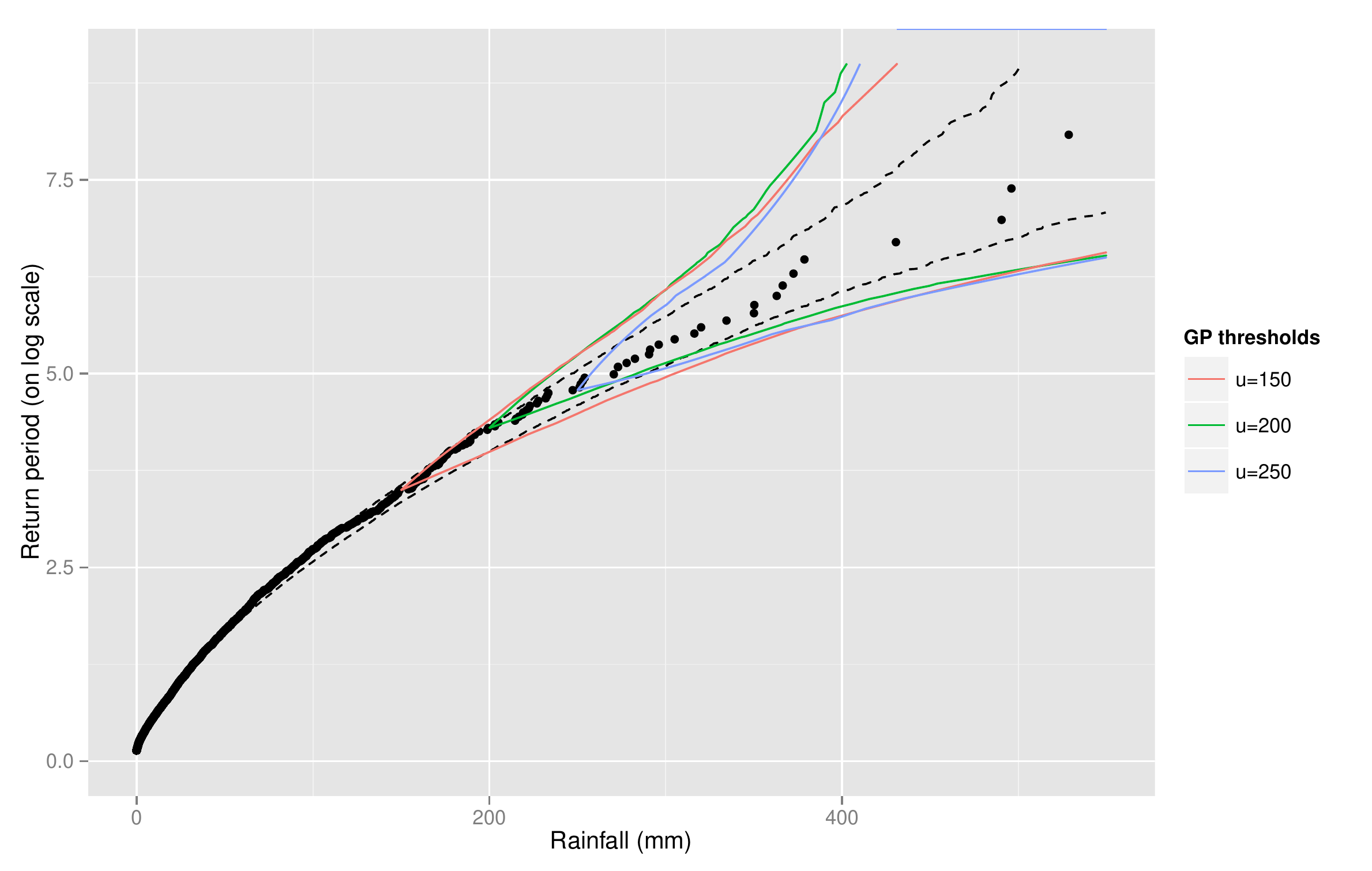}}
  \end{minipage}
  \begin{minipage}[b]{0.75\linewidth}
    \centering
   \centerline{ \includegraphics[trim = 0mm 2mm 0mm 5mm, clip,width=\textwidth]{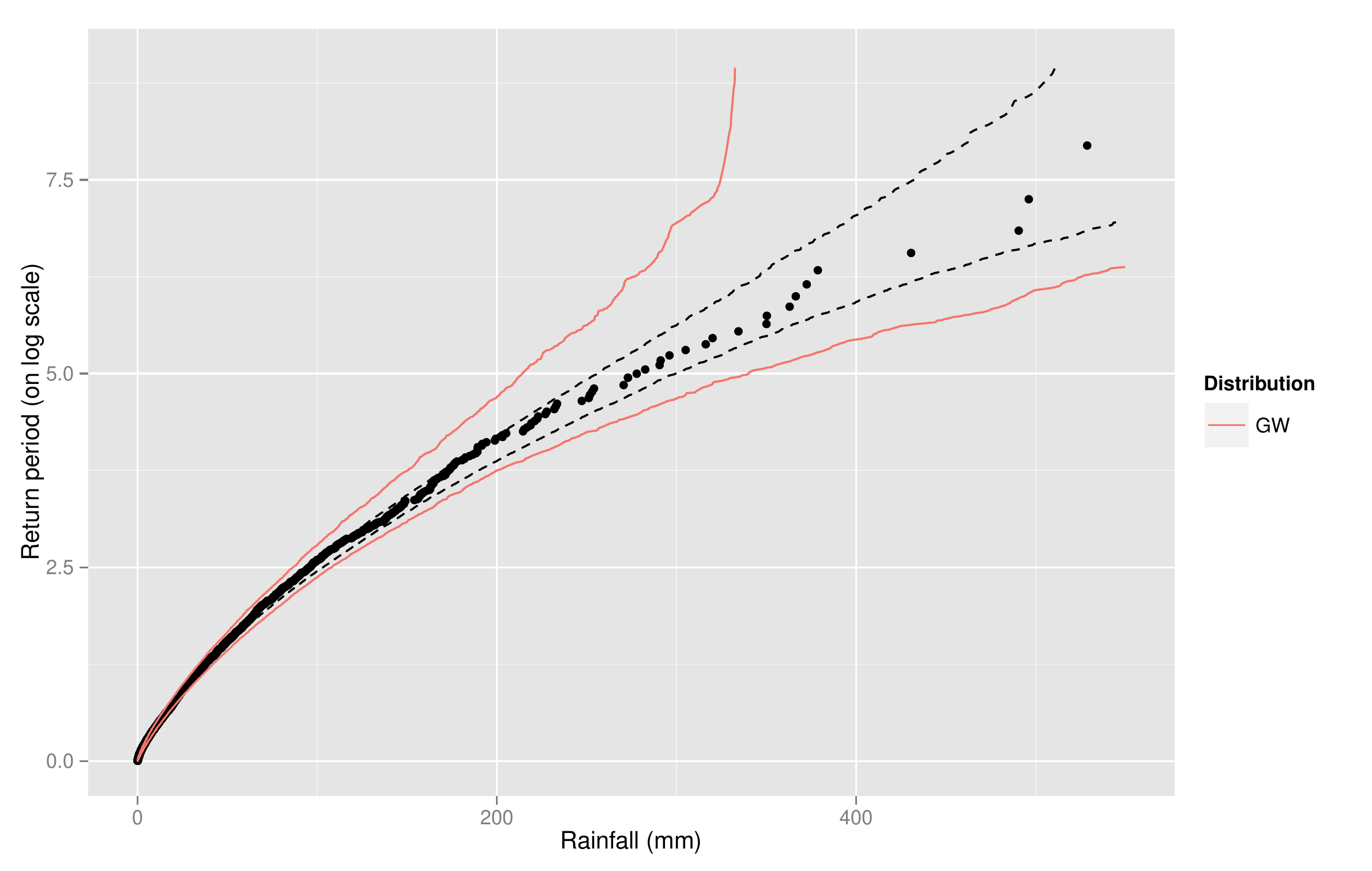}}
 \end{minipage}
 \caption{\footnotesize Return periods of the extreme winter weekly rainfall at Brisbane station. The broken curves represent the 95\% CI estimated by the ARMA-GARCH-GH model. In the left panel, the solid curves are the 95\% CIs produced by the Pareto model with different thresholds. In the right panel, the solid curves are the 95\% CIs produced by the generalised Weibull distribution.}
\label{comp}
\end{figure}
\indent Similar observations can be made when we compare the CIs for return periods produced by the GH and the GW model for the Brisbane weather station. It is clear that the generalised hyperbolic model has narrower CI for rainfall extremes. Furthermore, both the GP and GW methods focus solely on distributional properties of rainfall and rainfall extremes, and they fail to account for serial dependencies in rainfall data. In the ARMA-GARCH-GH model, however, the autocorrelation and heteroscedasticity are addressed by the ARMA and GARCH components so that short-term dependencies in the data can be efficiently captured. Thus it is suitable for evaluating duration over thresholds as well as low rainfall and dry periods. Similar results are observed for rainfall data from weather stations at the Oaks and Pardelup (Appendix B).

\section{Discussion}
The analysis presented in this paper demonstrates that the ARMA-GARCH-GH model has the potential to outperform conventional methods, such as the GP and the GW distribution for modelling rainfall and rainfall extremes. It appears that useful information from the centre of the data is being discarded for estimating extreme rainfall return periods by the GP approach. Our analysis also shows that the ARMA-GARCH-GH model not only fits the tails of the distribution, but fits the whole distribution well. The model uses an ARMA and GARCH structure to efficiently represents short-term dependencies in the data. This modelling approach is well suited for a number of applications where there is demand for return periods of extreme rainfall, durations of heavy and light rainfall, as well as dry periods.  In its comparison with the GP distribution, we see that the ARMA-GARCH-GH model produces much narrower CIs in the tails. Similarly it also outperformed the GW distribution in its ability to fit the entire rainfall distribution. Furthermore, the ARMA-GARCH-GH model allows us to evaluating duration over and below thresholds partly because it accounts for short-term dependencies in the data, and because it has an effective mechanism for dealing with zero rainfall through censoring.  \\
\indent There are various ways that the methods presented in this paper could be developed further to better fulfil application requirements. There has been a surge of interest in space-time modelling of rainfall and rainfall extremes. Hierarchical models based on well known time series modelling methods such as the dynamical linear models and the AR models are often used in the literature \citep{bakar2013sptimer,Sahu2010}. The development of a spatio-temporal version of the ARMA-GARCH-GH is currently being undertaken.

\section*{Acknowledgements}
The authors wish to acknowledge funding for this research from the Digital Productivity and Services Flagship, CSIRO. This work was also financially supported by AusAID. The authors would also like to thank Warren Jin, Steven Crimp and Kristen Williams who provided insightful comments on this paper.
\appendix
\section{Proofs of propositions}
\subsection{Proposition \ref{thm1}}
\label{app1}
\begin{proof}
We want to show that the GH distribution can match the tail decay of the generalised Pareto distribution given by
\begin{align}
\label{gp}
\lim_{x\to\infty}d_{GP}(x)\sim x^{-\left(\frac{1}{\kappa}+1\right)}, \kappa\ne0.
\end{align}
We first fix $\Sigma=1$ in the GH parameterisation to avoid the identification problem. To get some intuition on the proof, let us consider the reparametrisation
\begin{align}
\eta=\rho\xi,\>\>\>\xi=\frac{1}{\sqrt{1+\zeta}},\>{\text{\normalfont where}}\>\>\>\rho=\frac{\gamma}{\sqrt{\psi+\gamma^2}},\>\>\>\zeta=\sqrt{\chi\psi}.\notag
\end{align}
The $\left(\eta,\xi\right)$ parameterisation provides a useful view of the shape of the GH distribution. The parameter restrictions imply $0<\left\vert\eta\right\vert<\xi<1$, so all possible values for $\eta$ and $\xi$ lie in the interior of a triangle with corners $\left(-1,1\right)$, $\left(1,1\right)$ and $\left(0,0\right)$. This is the hyperbolic {\it shape triangle}. As it approaches the right hand side of the shape triangle, the distribution becomes more right-skewed with heavier tails. Thus we borrow the ideas of \cite{eberlein2004generalized} and approximate the right hand side boundary of the shape triangle, then we will investigate the properties of this limiting distribution as $x\to\infty$. One obvious way to achieve this is to let $\sqrt{\psi+\gamma^2}\to\infty$ and $\chi\to0$. We also assume
\begin{align}
\label{assumption1}
&\sqrt{\psi+\gamma^2}-\gamma=\delta>0,\\
\label{assumption2}
&\chi\sqrt{\psi+\gamma^2}=\tau>0.
\end{align}
To do this, we need to make use of several asymptotic properties of the Bessel function $K_v$, we write them down here explicitly:
\begin{align}
\label{B1}
&K_v(x)\sim\frac{1}{2}\Gamma(v)\left(\frac{x}{2}\right)^{-v},\>{\text{\normalfont for}}\>\>\>v>0, x\to0,\\
\label{B2}
&K_v(x)\sim\frac{1}{2}\Gamma(-v)\left(\frac{x}{2}\right)^{v},\>{\text{\normalfont for}}\>\>\>v<0, x\to0,\\
\label{B3}
&K_v(x)\sim\sqrt{\frac{\pi}{2x}}e^{-x},\>\>\>x\to\infty,\\
\label{B4}
&K_v(x)\sim-\ln\left(x\right),\>{\text{\normalfont for}}\>\>\>v\to0.
\end{align}
Using \eqref{B3}, for $\sqrt{\psi+\gamma^2}$ large enough, we have
\begin{align}
\frac{K_{\lambda-\frac{1}{2}}\left(\sqrt{\left(\chi+\left(x-\mu\right)^2\right)\left(\psi+\gamma^2\right)}\right)}{\sqrt{2\pi}\left(\sqrt{\left(\chi+\left(x-\mu\right)^2\right)\left(\psi+\gamma^2\right)}\right)^{\frac{1}{2}-\lambda}}\sim
\frac{\left(\sqrt{\left(\chi+\left(x-\mu\right)^2\right)\left(\psi+\gamma^2\right)}\right)^{\lambda-1}}{2\exp{\left(\sqrt{\left(\chi+\left(x-\mu\right)^2\right)\left(\psi+\gamma^2\right)}\right)}}.
\end{align}
Then
\begin{align}
d_{GH}(x)&\sim\frac{\left(\sqrt{\frac{\psi}{\chi}}\right)^\lambda\left(\psi+\gamma^2\right)^{\frac{1}{2}-\lambda}\exp{\left(\gamma\left(x-\mu\right)-\sqrt{\left(\chi+\left(x-\mu\right)^2\right)\left(\psi+\gamma^2\right)}\right)}}{2K_\lambda\left(\sqrt{\chi\psi}\right)\left(\sqrt{\left(\chi+\left(x-\mu\right)^2\right)\left(\psi+\gamma^2\right)}\right)^{1-\lambda}}\notag\\
\label{step0}
&=\frac{\psi^{\lambda/2}\left(\chi^{-\lambda/2}\left(\psi+\gamma^2\right)^{-\lambda/4}\right)\exp{\left(\gamma\left(x-\mu\right)-\sqrt{\left(\chi+\left(x-\mu\right)^2\right)\left(\psi+\gamma^2\right)}\right)}}{\left(\psi+\gamma^2\right)^{\lambda/4}2K_\lambda\left(\sqrt{\chi\psi}\right)\left(\chi+\left(x-\mu\right)^2\right)^{\left(1-\lambda\right)/2}}\\
\label{step}
&\sim\frac{\left(\frac{2\delta}{\tau}\right)^{\lambda/2}\exp{\left(\gamma\left(x-\mu\right)-\sqrt{\left(\psi+\gamma^2\right)}\sqrt{\left(\chi+\left(x-\mu\right)^2\right)}\right)}}{2K_\lambda\left(\sqrt{\chi\psi}\right)\left(\chi+\left(x-\mu\right)^2\right)^{\left(1-\lambda\right)/2}}\\
\label{step2}
&\sim\frac{\left(\frac{2\delta}{\tau}\right)^{\lambda/2}\exp{\left(\gamma\left(x-\mu\right)-\sqrt{\left(\psi+\gamma^2\right)}\left\vert x-\mu\right\vert\right)}}{2K_\lambda\left(\sqrt{2\delta\tau}\right)\left\vert x-\mu\right\vert^{1-\lambda}}.
\end{align}
From \eqref{step0} to \eqref{step} we used, for $\sqrt{\psi+\gamma^2}\to\infty$
\begin{align}
\frac{\psi^{\lambda/2}}{\left(\psi+\gamma^2\right)^{\lambda/4}}&=\frac{\left(\left(\sqrt{\psi+\gamma^2}\right)^2-\gamma^2\right)^{\lambda/2}}{\sqrt{\psi+\gamma^2}^{\lambda/2}}=\frac{\left(\left(\sqrt{\psi+\gamma^2}\right)^2-\left(\sqrt{\psi+\gamma^2}-\delta\right)^2\right)^{\lambda/2}}{\sqrt{\psi+\gamma^2}^{\lambda/2}}\notag\\
&=\left(2\delta-\frac{\delta^2}{\sqrt{\psi+\gamma^2}}\right)^{\lambda/2}\to\left(2\delta\right)^{\lambda/2}.
\end{align}
From \eqref{step} to \eqref{step2} we used
\begin{align}
K_\lambda\left(\sqrt{\chi\psi}\right)&=K_\lambda\left(\sqrt{\chi}\sqrt{\left(\sqrt{\psi+\gamma^2}\right)^2-\gamma^2}\right)=K_\lambda\left(\sqrt{\chi}\sqrt{\left(\sqrt{\psi+\gamma^2}\right)^2-\left(\sqrt{\psi+\gamma^2}-\delta\right)^2}\right)\notag\\
&=K_\lambda\left(\sqrt{2\delta\chi\sqrt{\psi+\gamma^2}-\chi\delta^2}\right)\to K_\lambda\left(\sqrt{2\delta\tau}\right),\>\>\>{\text{\normalfont for}}\>\chi\to0.
\end{align}
Using \eqref{assumption1}, it follows from \eqref{step2} that, for $x>\mu$
\begin{align}
\label{ghlimit}
d_{GH}(x)\sim\frac{\left(\frac{2\delta}{\tau}\right)^{\lambda/2}\exp{\left(-\delta\left(x-\mu\right)\right)}\left(x-\mu\right)^{\lambda-1}}{2 K_\lambda\left(\sqrt{\delta\tau}\right) }.
\end{align}
To finally obtain the limiting distribution as $\delta\to0$, we consider the following cases.\\
{\bf Case 1.} For $\lambda\in\mathbb{R}_-$. Using \eqref{B2}, \eqref{ghlimit} can be rewritten as
\begin{align}
\label{case1}
d_{GH}(x)&\sim \frac{\left(\frac{2\delta}{\tau}\right)^{\lambda/2}\left(x-\mu\right)^{\lambda-1}\exp{\left(-\delta \left(x-\mu\right)\right)}}{\Gamma\left(-\lambda\right)}\left(\frac{2}{\sqrt{\delta\tau}}\right)^\lambda=\frac{\left(\frac{4}{\tau}\right)^{\lambda}\left(x-\mu\right)^{\lambda-1}\exp{\left(-\delta \left(x-\mu\right)\right)}}{\Gamma\left(-\lambda\right)}.
\end{align}
Consequently
\begin{align}
\label{limcase1}
\lim_{\delta\to0}d_{GH}(x)\sim\frac{\left(\frac{4}{\tau}\right)^{\lambda}\left(x-\mu\right)^{\lambda-1}}{\Gamma\left(-\lambda\right)},
\end{align}
which is well-defined for $\lambda\nrightarrow-\infty$. Now, let $x\to\infty$ to get
\begin{align}
\lim_{x\to\infty}\lim_{\delta\to0}d_{GH}(x)\sim\frac{\left(\frac{4}{\tau}\right)^{\lambda}x^{\lambda-1}}{\Gamma\left(-\lambda\right)}.
\end{align}
To match its tail decay with \eqref{gp}, we only require $\lambda=-\frac{1}{\kappa}$ for $\kappa\in\mathbb{R}_+$.\\
{\bf Case 2.} For $\lambda=0$. Using \eqref{B4}, we have for $\delta\to0$
\begin{align}
 K_\lambda\left(\sqrt{\delta\tau}\right)\to-\ln\left(\sqrt{\delta\tau}\right)\to\infty,
\end{align}
so $\lim_{\delta\to\infty}d_{GH}(x)\to0$, and there is no limiting probability distribution. We should exclude this case.\\
{\bf Case 3.} For $\lambda\in\mathbb{R}_+$. Using \eqref{B1}, we have for $\delta\to0$
\begin{align}
\label{case3eq}
d_{GH}(x)\sim \frac{\left(\frac{2\delta}{\tau}\right)^{\lambda/2}x^{\lambda-1}\exp{\left(-\delta x\right)}}{\Gamma\left(\lambda\right)}\left(\frac{\sqrt{\delta\tau}}{2}\right)^\lambda=\frac{\left(\frac{\delta}{\sqrt{2}}\right)^{\lambda}x^{\lambda-1}\exp{\left(-\delta x\right)}}{\Gamma\left(\lambda\right)}\to0.
\end{align}
Hence there is no limiting distribution. However, it is well-known that generalised Pareto distirbutions converge to the exponential function as $\kappa\to0$, i.e.
\begin{align}
\label{wk}
\lim_{x\to\infty}\lim_{\kappa\to0}d_{GP(\kappa,\mu,\sigma)}(x)\sim\frac{1}{\sigma}\exp\left(-\frac{x}{\sigma}\right).
\end{align}
In this case, we do not need the limiting distribution as $\delta\to0$. The exponential decay of \eqref{wk} can be easily achieved by a hyperbolic distribution (i.e. $\lambda=1$), we have
\begin{align}
\lim_{x\to\infty}d_{GH}(x)\sim\frac{\left(\frac{2\delta}{\tau}\right)^{1/2}\exp{\left(-\delta x\right)}}{2 K_\lambda\left(\sqrt{\delta\tau}\right) }.
\end{align}
\end{proof}
\subsection{Proposition \ref{prop1}}
\label{app2}
\begin{proof}
We will make use of mean and variance expressions of the generalised inverse Gaussian distirbution, whose density function is given by
\begin{align}
\label{gigden}
d_{GIG\left(\lambda,\chi,\psi\right)}\left(x\right)=\left(\frac{\sqrt{\psi}}{\sqrt{\chi}}\right)^\lambda\frac{1}{2K_\lambda\left(\sqrt{\chi\psi}\right)}x^{\lambda-1}\exp{\left(-\frac{1}{2}\left(\frac{\chi}{x}+\psi x\right)\right)},\>\>\>x>0,
\end{align}
where $\chi,\psi>0$ and $\lambda\in\mathbb{R}$. It has the moment generating function
\begin{align}
\label{momg}
E\left(X^k\right)=\left(\frac{\sqrt{\chi}}{\sqrt{\psi}}\right)^k\frac{K_{\lambda+k}\left(\sqrt{\chi\psi}\right)}{K_{\lambda}\left(\sqrt{\chi\psi}\right)}.
\end{align}
Now for $X\sim GH(\lambda,\chi,\psi,\mu,\Sigma,\gamma)$, it follows from \eqref{mv} and \eqref{momg} that if we impose $Var(X)=1$, we then get
\begin{align}
\label{eq}
\frac{\gamma^2}{\psi}\left(\frac{K_{\lambda+2}\left(\sqrt{\psi}\right)}{K_{\lambda}\left(\sqrt{\psi}\right)}-\frac{K_{\lambda+1}^2\left(\sqrt{\psi}\right)}{K_{\lambda}^2\left(\sqrt{\psi}\right)}\right)+\frac{\Sigma}{\sqrt{\psi}}\frac{K_{\lambda+1}\left(\sqrt{\psi}\right)}{K_{\lambda}\left(\sqrt{\psi}\right)}=1,
\end{align}
Solving this simple linear equation for $\Sigma$, we have
\begin{align}
\label{muvar2}
\Sigma=\frac{\sqrt{\psi}}{M_{\lambda+1}\left(\sqrt{\psi}\right)}-\frac{\gamma^2}{\psi}\left(M_{\lambda+2}\left(\sqrt{\psi}\right)-M_{\lambda+1}\left(\sqrt{\psi}\right)\right),
\end{align}
where $M_{\lambda+1}\left(\sqrt{\psi}\right)=K_{\lambda+1}\left(\sqrt{\psi}\right)/K_{\lambda}\left(\sqrt{\psi}\right)$. However we notice that the expression for $\Sigma$ is not always postive. To overcome this problem, we replace the parameter $\gamma$ in \eqref{newformulation} with the parameter $\beta=\gamma\Sigma^{-1}$ to obtain the new parameterisation $X\sim GH(\lambda,\chi,\psi,\mu,\Sigma,\beta)$. Then \eqref{eq} becomes
\begin{align}
\label{eq2}
\Sigma^2\frac{\beta^2}{\psi}\left(\frac{K_{\lambda+2}\left(\sqrt{\psi}\right)}{K_{\lambda}\left(\sqrt{\psi}\right)}-\frac{K_{\lambda+1}^2\left(\sqrt{\psi}\right)}{K_{\lambda}^2\left(\sqrt{\psi}\right)}\right)+\frac{\Sigma}{\sqrt{\psi}}\frac{K_{\lambda+1}\left(\sqrt{\psi}\right)}{K_{\lambda}\left(\sqrt{\psi}\right)}-1=0,
\end{align}
whose solutions are
\begin{align}
\Sigma=\frac{-1\pm\sqrt{1+4\left(N_{\lambda+2}\left(\sqrt{\psi}\right)-1\right)\beta^2}}{2M_{\lambda+1}\left(\sqrt{\psi}\right)\left(N_{\lambda+2}\left(\sqrt{\psi}\right)-1\right)\beta^2/\sqrt{\psi}},
\end{align}
where $N_{\lambda+2}\left(\sqrt{\psi}\right)=K_{\lambda+2}\left(\sqrt{\psi}\right)K_{\lambda}\left(\sqrt{\psi}\right)/K_{\lambda+1}^2\left(\sqrt{\psi}\right)$.\\
\indent We know that $M_{v}\left(x\right)>0$ for all $v$ and $x>0$, and \cite{ismail1978monotonicity} showed that $N_{v}\left(x\right)>1$ for $\lambda\in\mathbb{R}$ and $x>0$, so \eqref{eq2} always has a postive and a negative solutions. Since $\Sigma>0$, we need to choose
\begin{align}
\Sigma=\frac{-1+\sqrt{1+4\left(N_{\lambda+2}\left(\sqrt{\psi}\right)-1\right)\beta^2}}{2M_{\lambda+1}\left(\sqrt{\psi}\right)\left(N_{\lambda+2}\left(\sqrt{\psi}\right)-1\right)\beta^2/\sqrt{\psi}}\>\>\>{\text{\normalfont and}}\>\>\>\mu=-\frac{\beta\Sigma M_{\lambda+1}\left(\sqrt{\psi}\right)}{\sqrt{\psi}}.\notag
\end{align}
\end{proof}

\section{Modelling results for the Oaks and Pardelup}

\begin{figure}[h]
 \centerline{\includegraphics[trim = 0mm 0mm 0mm 6mm, clip,width=150mm]{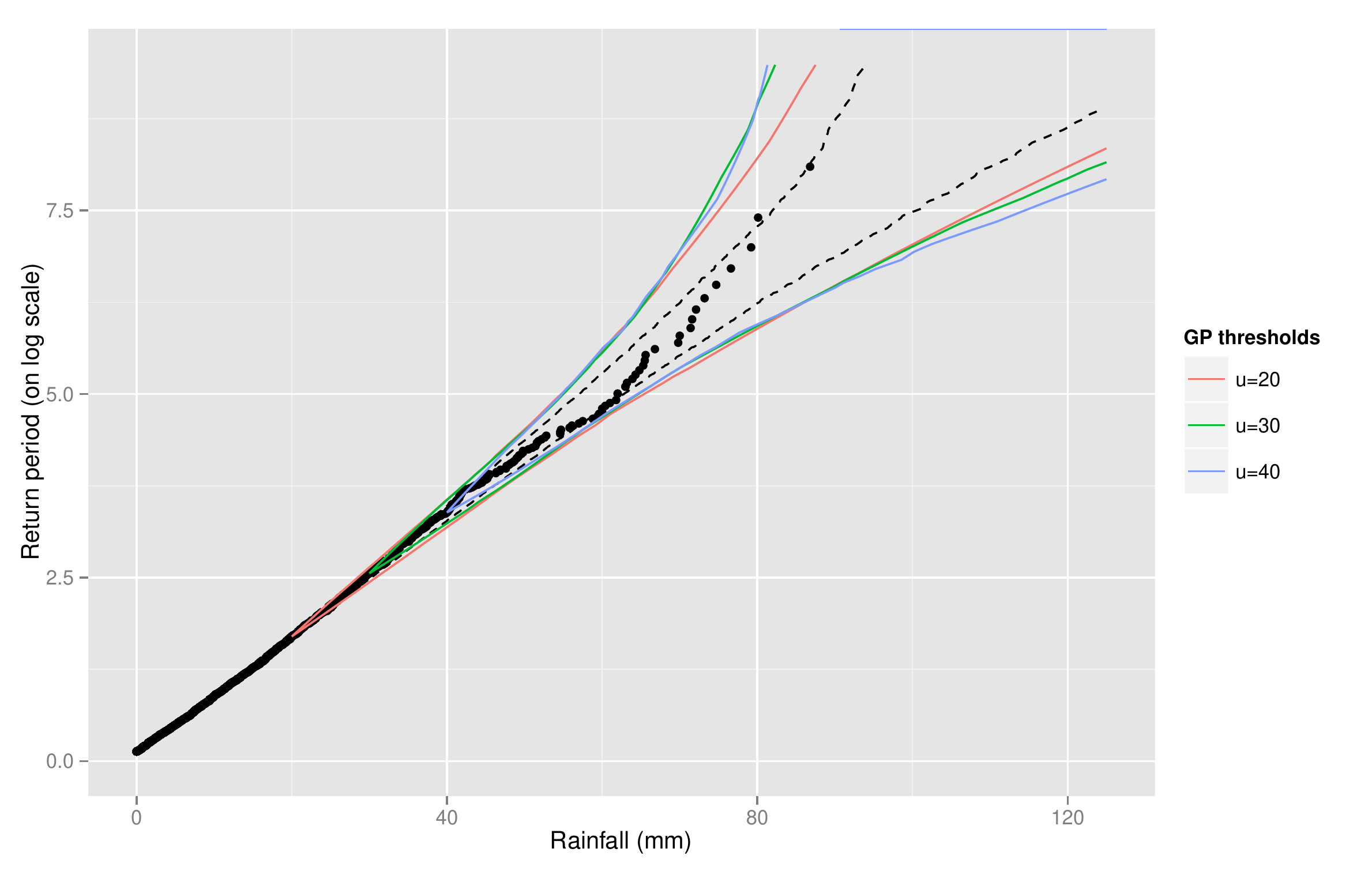}}
 \caption{\footnotesize Return periods of winter extreme weekly rainfall at the Oaks weather station, based on 1,000 simulations. The broken curves represent the 95\% CI estimated by the ARMA-GARCH-GH model, and the solid curves are the 95\% CIs produced by the Pareto model with different thresholds.}
\end{figure}

\begin{figure}[h]
 \centerline{\includegraphics[width=150mm]{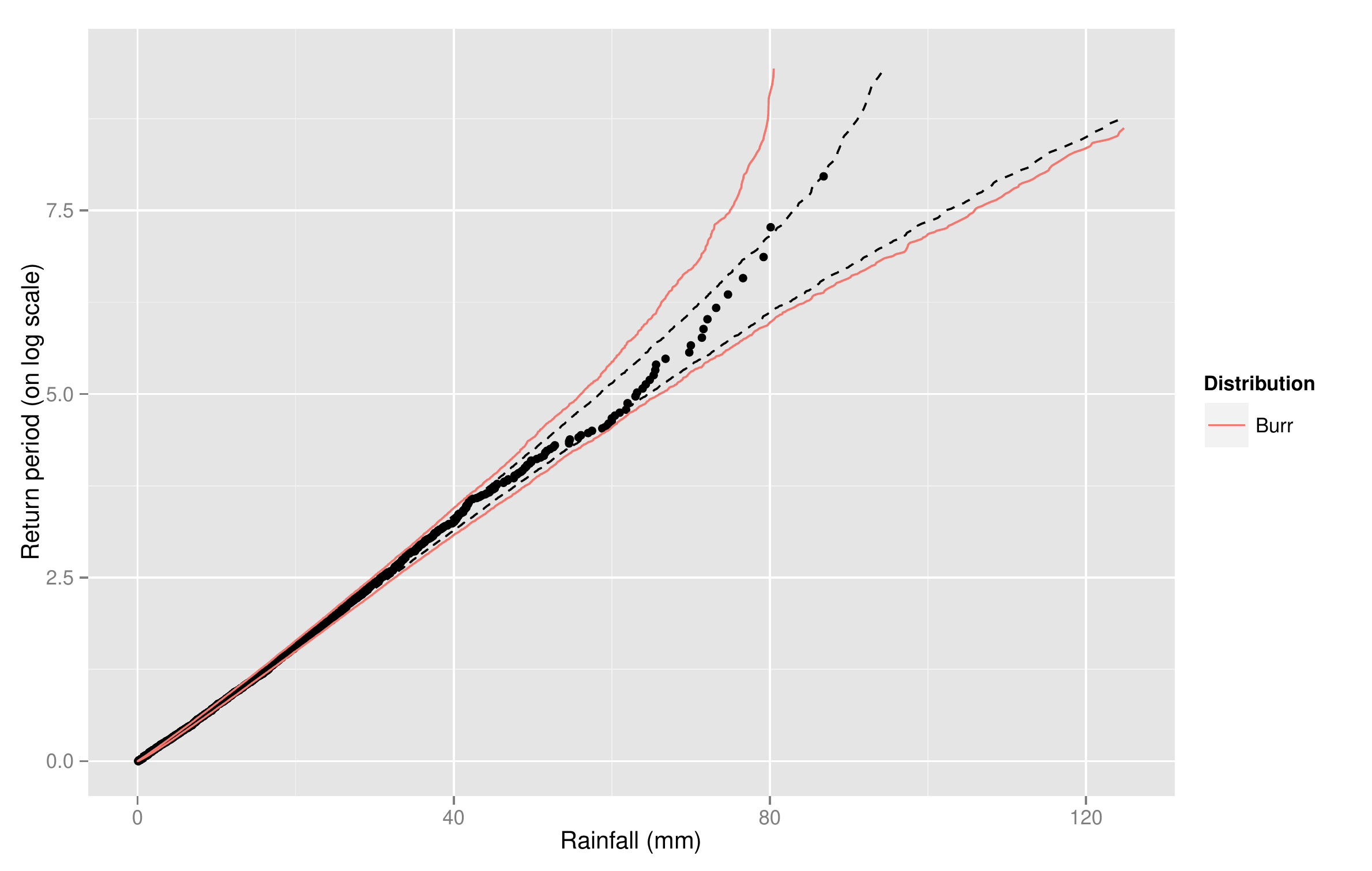}}
 \caption{\footnotesize Return periods of winter extreme weekly rainfall at the Oaks weather station, based on 1,000 simulations. The broken curves represent the 95\% CI estimated by the ARMA-GARCH-GH model, and the solid curves are the 95\% CIs produced by the Burr distribution.}
\end{figure}

\begin{figure}[h]
 \centerline{\includegraphics[width=190mm]{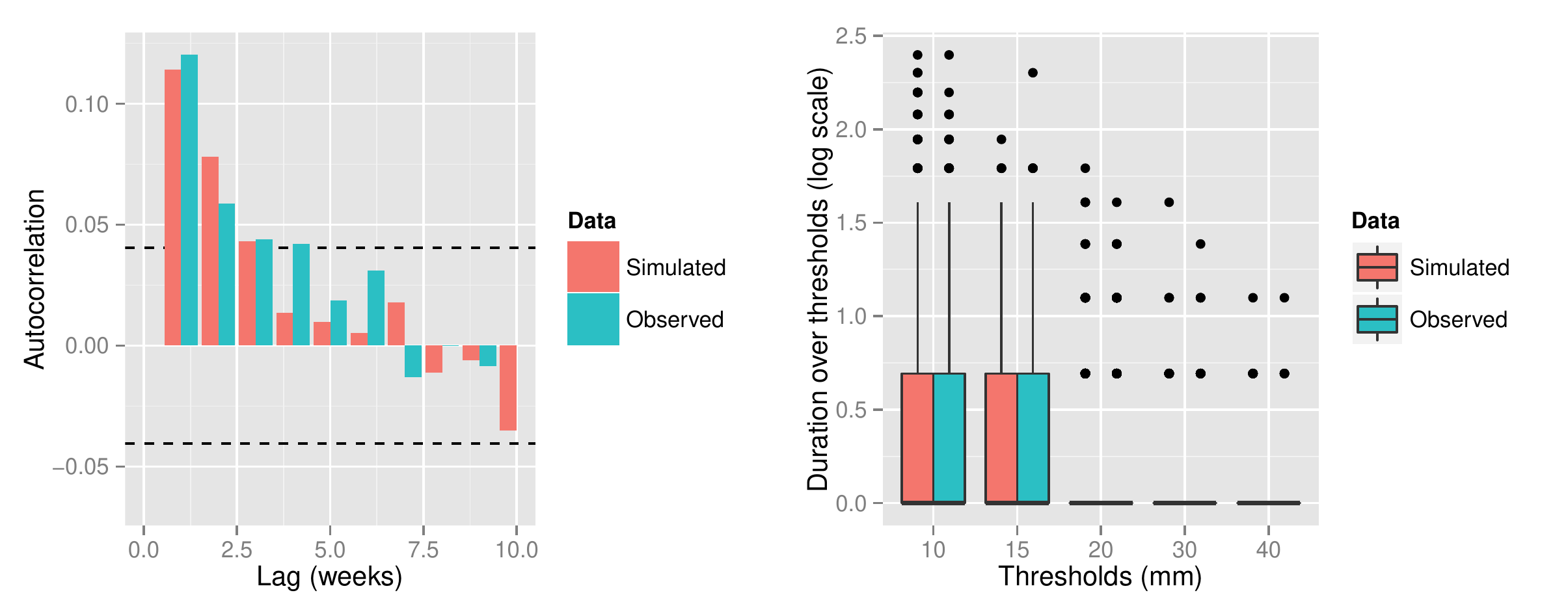}}
 \caption{\footnotesize The left panel shows the comparison of autocorrelation in the Oaks data and the data simulated by the ARMA-GARCH-GH model; the right panel is a boxplot (on log scale) of numbers of consecutive weeks with total rainfall over thresholds. Both are based on 1,000 simulations.}
\end{figure}

\begin{figure}[h]
 \centerline{\includegraphics[trim = 0mm 0mm 0mm 6mm, clip,width=150mm]{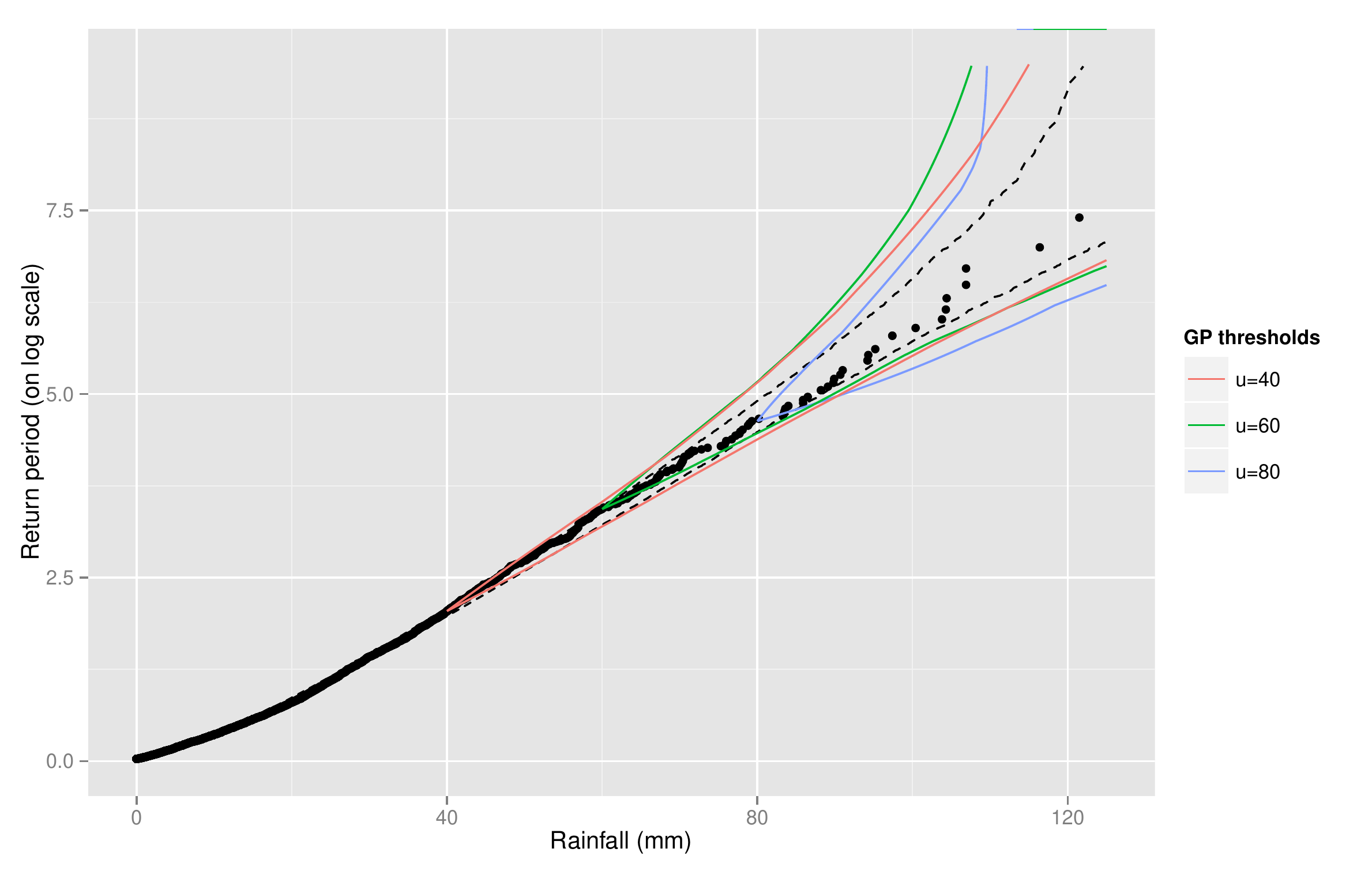}}
 \caption{\footnotesize Return periods of winter extreme weekly rainfall at Pardelup weather station, based on 1,000 simulations. The broken curves represent the 95\% CI estimated by the ARMA-GARCH-GH model, and the solid curves are the 95\% CIs produced by the Pareto model with different thresholds.}
\end{figure}

\begin{figure}[h]
 \centerline{\includegraphics[width=150mm]{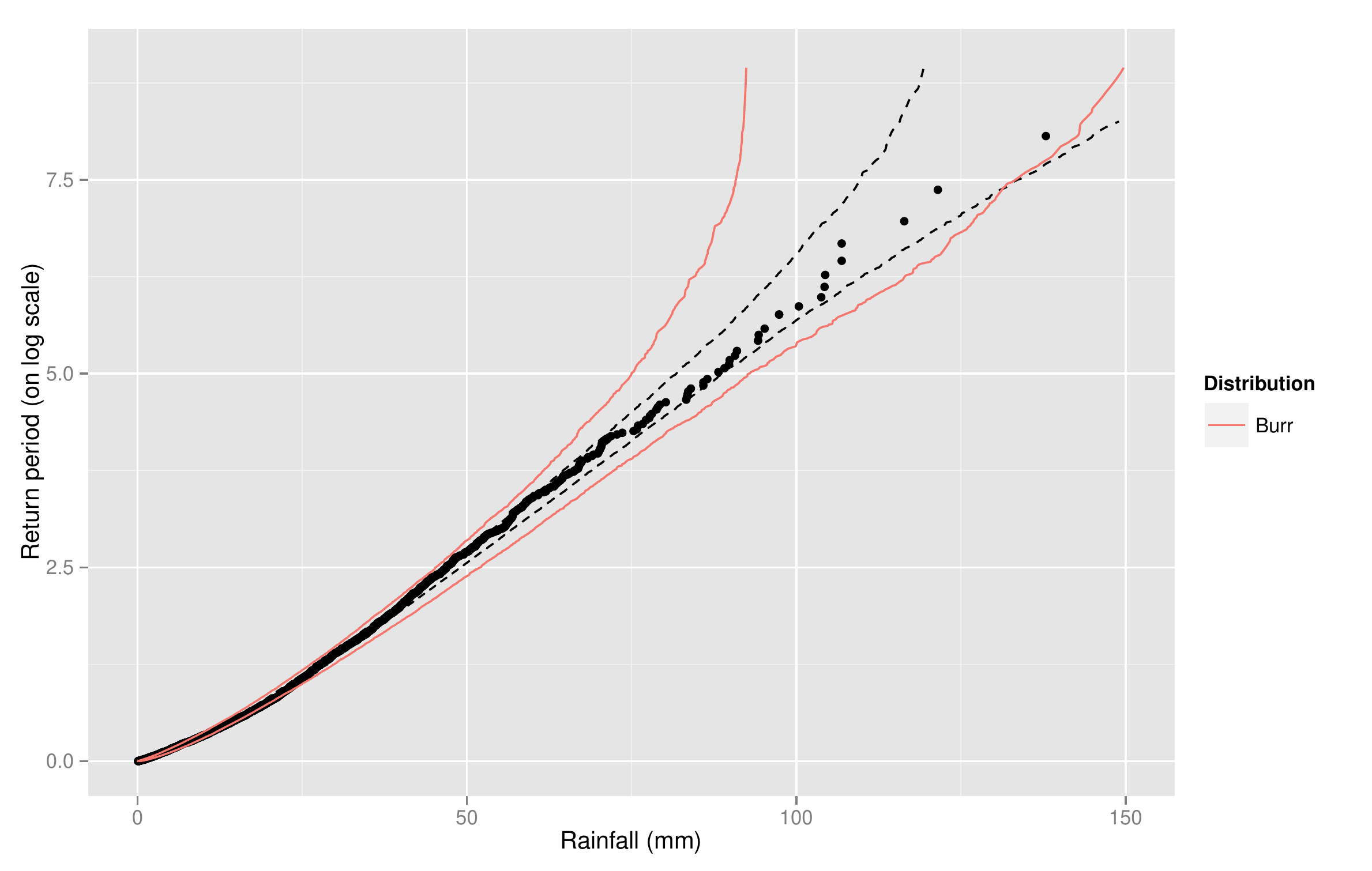}}
 \caption{\footnotesize Return periods of winter extreme weekly rainfall at Pardelup weather station, based on 1,000 simulations. The broken curves represent the 95\% CI estimated by the ARMA-GARCH-GH model, and the solid curves are the 95\% CIs produced by the Burr distribution.}
\end{figure}

\begin{figure}[h]
 \centerline{\includegraphics[width=190mm]{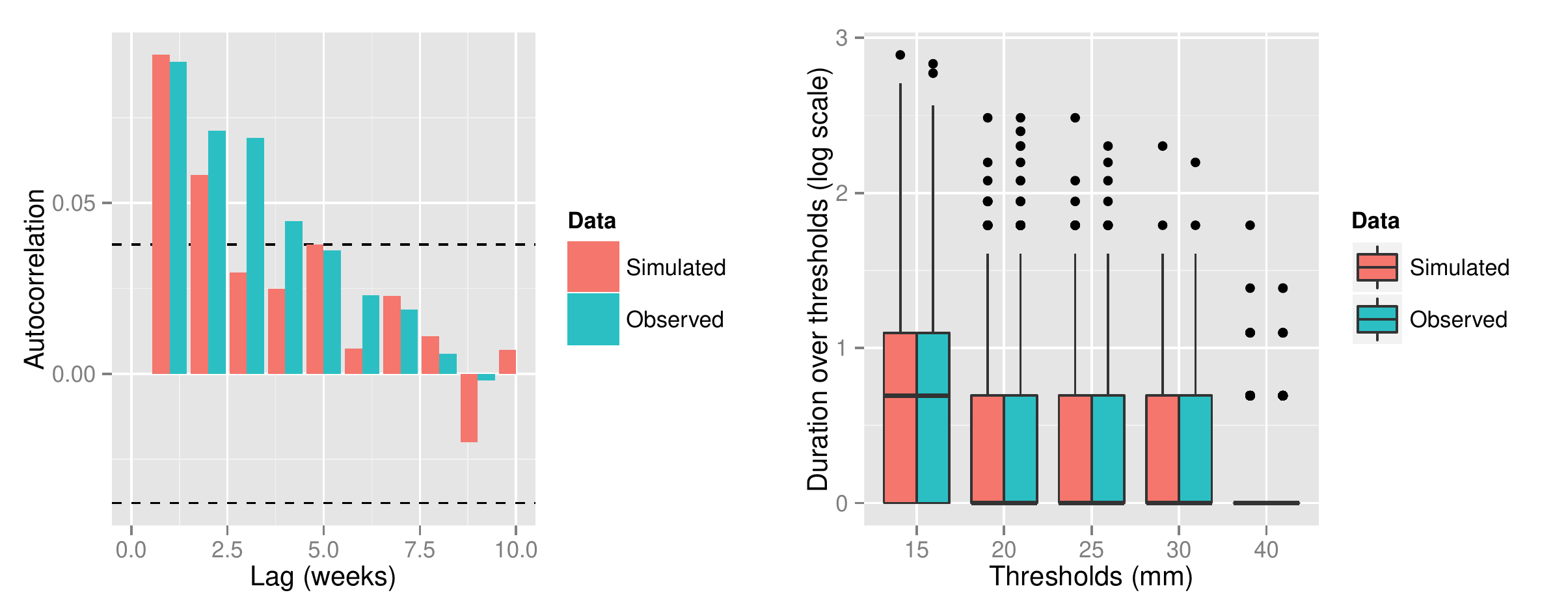}}
 \caption{\footnotesize The left panel shows the comparison of autocorrelation in Pardelup data and the data simulated by the ARMA-GARCH-GH model; the right panel is a boxplot (on log scale) of numbers of consecutive weeks with total rainfall over thresholds. Both are based on 1,000 simulations.}
\end{figure}

\bibliographystyle{Chicago}
\bibliography{GH}

\begin{thebibliography}{}

\bibitem[\protect\citeauthoryear{Ashcroft, French, and Chisholm}{Ashcroft
  et~al.}{2011}]{Ashcroft2011}
Ashcroft, M., K.~French, and L.~Chisholm (2011).
\newblock An evaluation of environmental factors affecting species
  distributions.
\newblock {\em Ecological Modelling\/}~{\em 222}, 524--531.

\bibitem[\protect\citeauthoryear{Ashcroft and Gollan}{Ashcroft and
  Gollan}{2012}]{Ashcroft2012}
Ashcroft, M. and J.~Gollan (2012).
\newblock Fine-resolution (25 m) topoclimatic grids of near-surface (5 cm)
  extreme temperatures and humidities across various habitats in a large (200 ×
  300 km) and diverse region.
\newblock {\em International Journal of Climatology\/}~{\em 32}, 2134--2148.

\bibitem[\protect\citeauthoryear{Bakar and Kokic}{Bakar and Kokic}{2014}]{shu1}
Bakar, K.~S. and P.~Kokic (2014).
\newblock A spatially varying approach for precipitation modelling in
  south-western australia.

\bibitem[\protect\citeauthoryear{Bakar and Sahu}{Bakar and
  Sahu}{2013}]{bakar2013sptimer}
Bakar, K.~S. and S.~K. Sahu (2013).
\newblock sptimer: Spatio-temporal bayesian modelling using r.

\bibitem[\protect\citeauthoryear{Barndorff-Nielsen}{Barndorff-Nielsen}{1977}]{barndorff1977}
Barndorff-Nielsen, O. (1977).
\newblock Exponentially decreasing distributions for the logarithm of particle
  size.
\newblock {\em Proceedings of the Royal Society of London. A. Mathematical and
  Physical Sciences\/}~{\em 353\/}(1674), 401--419.

\bibitem[\protect\citeauthoryear{Barndorff-Nielsen and
  Shephard}{Barndorff-Nielsen and Shephard}{2001}]{barndorff2001normal}
Barndorff-Nielsen, O.~E. and N.~Shephard (2001).
\newblock {\em Normal modified stable processes}.
\newblock MaPhySto, Department of Mathematical Sciences, University of Aarhus.

\bibitem[\protect\citeauthoryear{Bollerslev}{Bollerslev}{1986}]{bollerslev1986generalized}
Bollerslev, T. (1986).
\newblock Generalized autoregressive conditional heteroskedasticity.
\newblock {\em Journal of econometrics\/}~{\em 31\/}(3), 307--327.

\bibitem[\protect\citeauthoryear{Breymann and L{\"u}thi}{Breymann and
  L{\"u}thi}{2013}]{breymann2013ghyp}
Breymann, W. and D.~L{\"u}thi (2013).
\newblock ghyp: A package on generalized hyperbolic distributions.

\bibitem[\protect\citeauthoryear{Burr}{Burr}{1942}]{burr1942cumulative}
Burr, I.~W. (1942).
\newblock Cumulative frequency functions.
\newblock {\em The Annals of Mathematical Statistics\/}~{\em 13\/}(2),
  215--232.

\bibitem[\protect\citeauthoryear{Champernowne}{Champernowne}{1952}]{champernowne1952graduation}
Champernowne, D.~G. (1952).
\newblock The graduation of income distributions.
\newblock {\em Econometrica: Journal of the Econometric Society\/}, 591--615.

\bibitem[\protect\citeauthoryear{Coles, Bawa, Trenner, and Dorazio}{Coles
  et~al.}{2001}]{coles2001introduction}
Coles, S., J.~Bawa, L.~Trenner, and P.~Dorazio (2001).
\newblock {\em An introduction to statistical modeling of extreme values},
  Volume 208.
\newblock Springer.

\bibitem[\protect\citeauthoryear{Coles, Pericchi, and Sisson}{Coles
  et~al.}{2003}]{coles2003fully}
Coles, S., L.~R. Pericchi, and S.~Sisson (2003).
\newblock A fully probabilistic approach to extreme rainfall modeling.
\newblock {\em Journal of Hydrology\/}~{\em 273\/}(1), 35--50.

\bibitem[\protect\citeauthoryear{Davison and Smith}{Davison and
  Smith}{1990}]{davison1990models}
Davison, A.~C. and R.~L. Smith (1990).
\newblock Models for exceedances over high thresholds.
\newblock {\em Journal of the Royal Statistical Society. Series B
  (Methodological)\/}, 393--442.

\bibitem[\protect\citeauthoryear{Drosdowsky and Chambers}{Drosdowsky and
  Chambers}{2001}]{drosdowsky2001near}
Drosdowsky, W. and L.~E. Chambers (2001).
\newblock Near-global sea surface temperature anomalies as predictors of
  australian seasonal rainfall.
\newblock {\em Journal of Climate\/}~{\em 14\/}(7), 1677--1687.

\bibitem[\protect\citeauthoryear{DuMouchel}{DuMouchel}{1983}]{dumouchel1983estimating}
DuMouchel, W.~H. (1983).
\newblock Estimating the stable index $\alpha$ in order to measure tail
  thickness: a critique.
\newblock {\em The Annals of Statistics\/}, 1019--1031.

\bibitem[\protect\citeauthoryear{Eberlein and Hammerstein}{Eberlein and
  Hammerstein}{2004}]{eberlein2004generalized}
Eberlein, E. and E.~A.~v. Hammerstein (2004).
\newblock Generalized hyperbolic and inverse gaussian distributions: limiting
  cases and approximation of processes.
\newblock In {\em Seminar on Stochastic Analysis, Random Fields and
  Applications IV}, pp.\  221--264. Springer.

\bibitem[\protect\citeauthoryear{Finkenstadt and Rootz{\'e}n}{Finkenstadt and
  Rootz{\'e}n}{2004}]{finkenstadt2004extreme}
Finkenstadt, B. and H.~Rootz{\'e}n (2004).
\newblock {\em Extreme values in finance, telecommunications, and the
  environment}.
\newblock CRC Press.

\bibitem[\protect\citeauthoryear{Fowler, Blenkinsop, and Tebaldi}{Fowler
  et~al.}{2007}]{Fowler2007}
Fowler, H.~J., S.~Blenkinsop, and C.~Tebaldi (2007).
\newblock Linking climate change modelling to impacts studies: recent advances
  in downscaling techniques for hydrological modelling.
\newblock {\em International Journal of Climatology\/}~{\em 27}, 1547–--1578.

\bibitem[\protect\citeauthoryear{Gilli et~al.}{Gilli
  et~al.}{2006}]{gilli2006application}
Gilli, M. et~al. (2006).
\newblock An application of extreme value theory for measuring financial risk.
\newblock {\em Computational Economics\/}~{\em 27\/}(2-3), 207--228.

\bibitem[\protect\citeauthoryear{Hastings}{Hastings}{1970}]{hastings}
Hastings, W.~K. (1970).
\newblock Monte carlo sampling methods using markov chains and their
  applications.
\newblock {\em Biometrika\/}~{\em 57\/}(1), 97--109.

\bibitem[\protect\citeauthoryear{Hosking and Wallis}{Hosking and
  Wallis}{1987}]{hosking1987parameter}
Hosking, J.~R. and J.~R. Wallis (1987).
\newblock Parameter and quantile estimation for the generalized pareto
  distribution.
\newblock {\em Technometrics\/}~{\em 29\/}(3), 339--349.

\bibitem[\protect\citeauthoryear{Hutchinson}{Hutchinson}{1995}]{Hutchinson1995}
Hutchinson, M. (1995).
\newblock Stochastic space-time weather models from ground-based data.
\newblock {\em Agricultural and Forest Meteorology\/}~{\em 73}, 237--264.

\bibitem[\protect\citeauthoryear{Ismail and Muldoon}{Ismail and
  Muldoon}{1978}]{ismail1978monotonicity}
Ismail, M.~E. and M.~E. Muldoon (1978).
\newblock Monotonicity of the zeros of a cross-product of bessel functions.
\newblock {\em SIAM Journal on Mathematical Analysis\/}~{\em 9\/}(4), 759--767.

\bibitem[\protect\citeauthoryear{Joe}{Joe}{1987}]{joe1987estimation}
Joe, H. (1987).
\newblock Estimation of quantiles of the maximum of n observations.
\newblock {\em Biometrika\/}~{\em 74\/}(2), 347--354.

\bibitem[\protect\citeauthoryear{Keating, Carberry, Hammer, Probert, Robertson,
  Holzworth, Huth, Hargreaves, Meinke, Hochman, and et~al.}{Keating
  et~al.}{2003}]{Keating2003}
Keating, B.~A., P.~S. Carberry, G.~L. Hammer, M.~E. Probert, M.~J. Robertson,
  D.~Holzworth, N.~I. Huth, J.~N. Hargreaves, H.~Meinke, Z.~Hochman, and et~al.
  (2003).
\newblock An overview of {APSIM}, a model designed for farming systems
  simulation.
\newblock {\em European Journal of Agronomy\/}~{\em 18}, 267–288.

\bibitem[\protect\citeauthoryear{King, Alexander, and Donat}{King
  et~al.}{2013}]{king2013asymmetry}
King, A.~D., L.~V. Alexander, and M.~G. Donat (2013).
\newblock Asymmetry in the response of eastern australia extreme rainfall to
  low-frequency pacific variability.
\newblock {\em Geophysical Research Letters\/}~{\em 40\/}(10), 2271--2277.

\bibitem[\protect\citeauthoryear{Kleiber and Kotz}{Kleiber and
  Kotz}{2003}]{kleiber2003statistical}
Kleiber, C. and S.~Kotz (2003).
\newblock {\em Statistical size distributions in economics and actuarial
  sciences}, Volume 470.
\newblock John Wiley \& Sons.

\bibitem[\protect\citeauthoryear{Kokic, Jin, and Crimp}{Kokic
  et~al.}{2013}]{Kokic2013}
Kokic, P., H.~Jin, and S.~Crimp (2013).
\newblock Improved point scale climate projections using a block bootstrap
  simulation and quantile matching method.
\newblock {\em Climate Dynamics\/}~{\em 41}, 853--866.

\bibitem[\protect\citeauthoryear{K.S.Bakar and Jin}{K.S.Bakar and
  Jin}{2014}]{spdy1}
K.S.Bakar, a.~P. and H.~Jin (2014).
\newblock A spatio-dynamic model for assessing frost risk in south-east
  australia.

\bibitem[\protect\citeauthoryear{Li, Cai, and Campbell}{Li
  et~al.}{2005}]{li2005statistical}
Li, Y., W.~Cai, and E.~Campbell (2005).
\newblock Statistical modeling of extreme rainfall in southwest western
  australia.
\newblock {\em Journal of Climate\/}~{\em 18\/}(6).

\bibitem[\protect\citeauthoryear{Maddala}{Maddala}{1986}]{maddala1986limited}
Maddala, G.~S. (1986).
\newblock {\em Limited-dependent and qualitative variables in econometrics}.
\newblock Number~3. Cambridge university press.

\bibitem[\protect\citeauthoryear{Mencia and Sentana}{Mencia and
  Sentana}{2004}]{JEd}
Mencia, J. and E.~Sentana (2004).
\newblock Estimation and testing of dynamic models with generalised hyperbolic
  innovations.

\bibitem[\protect\citeauthoryear{Metropolis, Rosenbluth, Rosenbluth, Teller,
  and Teller}{Metropolis et~al.}{1953}]{metropolis}
Metropolis, N., A.~W. Rosenbluth, M.~N. Rosenbluth, A.~H. Teller, and E.~Teller
  (1953).
\newblock Equation of state calculations by fast computing machines.
\newblock {\em The journal of chemical physics\/}~{\em 21\/}(6), 1087--1092.

\bibitem[\protect\citeauthoryear{Mudholkar and Huston}{Mudholkar and
  Huston}{1996}]{Mudholkar1996}
Mudholkar, G.~S. and A.~D. Huston (1996).
\newblock The exponentiated {W}eibull family: some properties and a flood data
  application.
\newblock {\em Communication in Statistics - Theory and Methods\/}~{\em 25},
  3059--3083.

\bibitem[\protect\citeauthoryear{Pickands~III}{Pickands~III}{1971}]{pickands1971two}
Pickands~III, J. (1971).
\newblock The two-dimensional poisson process and extremal processes.
\newblock {\em Journal of Applied Probability\/}, 745--756.

\bibitem[\protect\citeauthoryear{Sahu, Gelfand, and Holland}{Sahu
  et~al.}{2010}]{Sahu2010}
Sahu, S.~K., A.~E. Gelfand, and D.~M. Holland (2010).
\newblock Fusing point and areal level space--time data with application to wet
  deposition.
\newblock {\em Journal of the Royal Statistical Society: Series C (Applied
  Statistics)\/}~{\em 59}, 77--103.

\bibitem[\protect\citeauthoryear{Schepen, Wang, and Robertson}{Schepen
  et~al.}{2012}]{schepen2012evidence}
Schepen, A., Q.~Wang, and D.~Robertson (2012).
\newblock Evidence for using lagged climate indices to forecast australian
  seasonal rainfall.
\newblock {\em Journal of Climate\/}~{\em 25\/}(4), 1230--1246.

\bibitem[\protect\citeauthoryear{Shao, Wang, and Zhang}{Shao
  et~al.}{2013}]{Shao2013}
Shao, Q., Q.~Wang, and L.~Zhang (2013).
\newblock A stochastic weather generation method for temporal precipitation
  simulation.
\newblock In {\em 20th International Congress on Modelling and Simulation,
  Adelaide, Australia, 1–6 December, 2013}, pp.\  2681--2687.

\bibitem[\protect\citeauthoryear{Shao, Wong, Xia, and W.-C.}{Shao
  et~al.}{2004}]{Shao2004}
Shao, Q., H.~Wong, J.~Xia, and I.~W.-C. (2004).
\newblock Models for extremes using the extended three parameter {B}urr {XII}
  system with application to flood frequency analysis.
\newblock {\em Hydrological Sciences Journal\/}~{\em 49}, 685--702.

\bibitem[\protect\citeauthoryear{Smith}{Smith}{1984}]{smith1984threshold}
Smith, R.~L. (1984).
\newblock Threshold methods for sample extremes.
\newblock In {\em Statistical extremes and applications}, pp.\  621--638.
  Springer.

\bibitem[\protect\citeauthoryear{Smith}{Smith}{1987}]{smith1987estimating}
Smith, R.~L. (1987).
\newblock Estimating tails of probability distributions.
\newblock {\em The annals of Statistics\/}, 1174--1207.

\bibitem[\protect\citeauthoryear{Stone, Hammer, and Marcussen}{Stone
  et~al.}{1996}]{stone1996prediction}
Stone, R.~C., G.~L. Hammer, and T.~Marcussen (1996).
\newblock Prediction of global rainfall probabilities using phases of the
  southern oscillation index.

\bibitem[\protect\citeauthoryear{Tetzlaff, Uhlenbrook, and Molnar}{Tetzlaff
  et~al.}{2005}]{Tetzlaff2005}
Tetzlaff, D., S.~Uhlenbrook, and P.~Molnar (2005).
\newblock Significance of spatial variability in precipitation for
  process-oriented modelling: results from two nested catchments using radar
  and ground station data.
\newblock {\em Hydrology \& Earth System Sciences\/}~{\em 9}.

\bibitem[\protect\citeauthoryear{Van~Montfort and Witter}{Van~Montfort and
  Witter}{1986}]{van1986generalized}
Van~Montfort, M. and J.~Witter (1986).
\newblock The generalized pareto distribution applied to rainfall depths.
\newblock {\em Hydrological Sciences Journal\/}~{\em 31\/}(2), 151--162.

\bibitem[\protect\citeauthoryear{Wang, Shrestha, Robertson, and Pokhrel}{Wang
  et~al.}{2012}]{Wang2012}
Wang, Q.~J., D.~L. Shrestha, D.~E. Robertson, and P.~Pokhrel (2012).
\newblock A log-sinh transformation for data normalization and variance
  stabilization.
\newblock {\em Water Resources Research\/}~{\em 48}, 1--7.

\bibitem[\protect\citeauthoryear{Whittle}{Whittle}{1951}]{arma}
Whittle, P. (1951).
\newblock {\em Hypothesis Testing in Time Series Analysis}.
\newblock Ph.\ D. thesis, University of Uppsala.

\bibitem[\protect\citeauthoryear{Ye and Li}{Ye and Li}{2011}]{ye2011method}
Ye, W. and Y.~Li (2011).
\newblock A method of applying daily gcm outputs in assessing climate change
  impact on multiple day extreme precipitation for brisbane river catchment.
\newblock In {\em 19th international Congress on modelling and simulation},
  pp.\  12--16.

\bibitem[\protect\citeauthoryear{Yiou, Goubanova, Li, and Nogaj}{Yiou
  et~al.}{2008}]{yiou2008weather}
Yiou, P., K.~Goubanova, Z.~Li, and M.~Nogaj (2008).
\newblock Weather regime dependence of extreme value statistics for summer
  temperature and precipitation.
\newblock {\em Nonlinear Processes in Geophysics\/}~{\em 15\/}(3).

\end{thebibliography}
\end{document}